\documentclass{aastex61}
\usepackage{CJK}
\usepackage{natbib}

\newcommand{\kms}{km\,s$^{-1}$}

\begin{document}


\title{SiS in the circumstellar envelope of IRC +10216: maser and quasi-thermal emission}
\begin{CJK}{UTF8}{gbsn}
%% Use \author, \affil, plus the \and command to format author and affiliation 
%% information.  If done correctly the peer review system will be able to
%% automatically put the author and affiliation information from the manuscript
%% and save the corresponding author the trouble of entering it by hand.
%%
%% The \affil should be used to document primary affiliations and the
%% \altaffil should be used for secondary affiliations, titles, or email.

%% Authors with the same affiliation can be grouped in a single
\correspondingauthor{Y. Gong}
\email{ygong@pmo.ac.cn, gongyan2444@gmail.com}

\author{Y. Gong}
\affil{Purple Mountain Observatory \& Key Laboratory of Radio Astronomy, Chinese Academy of Sciences, 2 West Beijing Road, 210008 Nanjing, PR China}
\affil{Max-Planck Institut f\"ur Radioastronomie, Auf Dem H\"ugel 69, 53121 Bonn, Germany}

\author{C. Henkel}
\affil{Max-Planck Institut f\"ur Radioastronomie, Auf Dem H\"ugel 69, 53121 Bonn, Germany}
\affil{Astronomy Department, King Abdulaziz University, PO Box 80203, 21589 Jeddah, Saudi Arabia}

\author{J. Ott}
\affil{National Radio Astronomy Observatory, P.O. Box O, 1003 Lopezville Road, Socorro, NM 87801, USA}

\author{K.~M. Menten}
\affil{Max-Planck Institut f\"ur Radioastronomie, Auf Dem H\"ugel 69, 53121 Bonn, Germany}

\author{M.~R. Morris}
\affil{Department of Physics and Astronomy, University of California, Los Angeles, California 90095-1547, USA}

\author{D. Keller}
\affil{Max-Planck Institut f\"ur Radioastronomie, Auf Dem H\"ugel 69, 53121 Bonn, Germany}
\affil{Instituut voor Sterrenkunde, Katholieke Universiteit Leuven, Celestijnenlaan 200D, 3001 Leuven, Belgium}

\author{M.~J. Claussen}
\affil{National Radio Astronomy Observatory, P.O. Box O, 1003 Lopezville Road, Socorro, NM 87801, USA}

\author{M. Grasshoff}
\affil{Visiting the Max-Planck Institut f\"ur Radioastronomie, Auf Dem H\"ugel 69, 53121 Bonn, Germany}

\author{R.~Q. Mao (毛瑞青)}
\affil{Purple Mountain Observatory \& Key Laboratory of Radio Astronomy, Chinese Academy of Sciences, 2 West Beijing Road, 210008 Nanjing, PR China}

%\author{Y. Gong\altaffilmark{1, 2}, C. Henkel\altaffilmark{2, 3}, J. Ott\altaffilmark{4}, K.~M. Menten\altaffilmark{2}, M.~R. Morris\altaffilmark{5}, D. Keller\altaffilmark{2, 6}, M.~J. Claussen\altaffilmark{4}, M. Grasshoff\altaffilmark{7}, R.~Q. Mao\altaffilmark{1}}


\begin{abstract}
  We present new Effelsberg-100 m, ATCA, and VLA observations of rotational SiS transitions in the circumstellar envelope (CSE) of IRC +10216. Thanks to the high angular resolution achieved by the ATCA observations, we unambiguously confirm that the molecule's $J=1\to 0$ transition exhibits maser action in this CSE, as first suggested more than thirty years ago. The maser emission's radial velocity peaking at a local standard of rest velocity of $-$39.862$\pm$0.065~\kms\,indicates that it arises from an almost fully accelerated shell. Monitoring observations show time variability of the SiS (1$\to$0) maser. The two lowest-$J$ SiS quasi-thermal emission lines trace a much more extended emitting region than previous high-$J$ SiS observations. Their distributions show that the SiS quasi-thermal emission consists of two components: one is very compact (radius$<1\rlap{.}''5$, corresponding to $<$3$\times 10^{15}$ cm), and the other extends out to a radius $>$11\arcsec. An incomplete shell-like structure is found in the north-east, which is indicative of existing SiS shells. Clumpy structures are also revealed in this CSE. The gain of the SiS (1$\to$0) maser (optical depths of about $-$5 at the blue-shifted side and, assuming inversion throughout the entire line's velocity range, about $-$2 at the red-shifted side) suggests that it is unsaturated. The SiS (1$\to$0) maser can be explained in terms of ro-vibrational excitation caused by infrared pumping, and we propose that infrared continuum emission is the main pumping source.
\end{abstract}




\section{Introduction} \label{sec:intro}
IRC +10216 (CW Leonis) is one of the brightest infrared objects in the sky \citep{1969ApJ...158L.133B}, and is regarded as the archetypal asymptotic giant branch (AGB) C-rich star. Its elementary abundance ratio is [C]/[O]$>$1.4 in its atmosphere \citep[e.g.,][]{1970ApJ...162L..15H,1997A&A...317..503G}. The star has probably reached a very late stage of its AGB evolution, shortly before turning into a preplanetary nebula \citep[e.g.,][]{1998MNRAS.300L..29S,2000A&A...357..169O}.
Models of the CO emission have led to distance estimates of 110--150 pc \citep{1997ApJ...483..913C,1998A&A...338..491G}, and a distance of 130 pc is adopted in this paper. IRC +10216 exhibits an extremely high mass-loss rate of $2\times10^{-5}$ $M_{\odot}$~yr$^{-1}$ \citep[][scaled to a distance of 130 pc]{1997ApJ...483..913C}, creating a nearly spherical, dense, circumstellar envelope (CSE) in which more than 80 molecular species have already been detected. These molecules include very unusual long carbon chains, metal cyanides, metal halides, and all known interstellar anions \citep[e.g.,][]{1987A&A...183L..10C,2000A&AS..142..181C,2008ApJ...688L..83C,2015A&A...574A..56G}. In addition, IRC +10216 is not only a Mira variable star which shows periodic variability at infrared and radio bands \citep{1992A&AS...94..377L,2006A&A...453..301M,1998ApJ...502..833M,2012A&A...543A..73M,2012ApJ...744..133M}, but it also displays time variations in the intensity of molecular lines at submillimeter and far-infrared wavelengths \citep{2014ApJ...796L..21C}. These properties make it an exceptional source for detailed investigations of physical and chemical processes related to C-rich AGB stars.

Masers in CSEs of evolved stars are powerful tools to constrain their physical parameters \citep[e.g.,][]{1996A&ARv...7...97H,2014ARA&A..52..339R}. Owing to different physical and chemical environments, CSEs of C-rich ([C]/[O]$>$1) and O-rich ([C]/[O]$<$1) stars show different kinds of masers. OH, SiO, and H$_{2}$O masers are known to be very common in CSEs of O-rich stars, but these masers are not detected toward C-rich stars \citep[e.g.,][]{2007IAUS..242..471H}. Nevertheless, there are a few transitions of HCN and SiS which have been proposed to be masers in the CSE of the extremely C-rich star IRC +10216 \citep[e.g.,][]{1981A&A...101..238G,1983ApJ...267..184H,2000ApJ...528L..37S,2003ApJ...583..446S,2006ApJ...646L.127F}. Among them, SiS (1$\to$0) was first proposed to be a maser more than 30 years ago \citep{1981A&A...101..238G,1983ApJ...267..184H}. Follow-up monitoring studies of this line suggested that the line was likely time-variable \citep{1984ApJ...286..276N,1985A&A...147..143H}. However, the SiS (1$\to$0) line of IRC +10216 has only been measured with $\ge$50\arcsec\,angular resolution so far, and the conclusion that the SiS (1$\to$0) line is a maser was tentative. Therefore, the maser nature of SiS (1$\to$0) needs to be confirmed with follow-up observations.

The molecule SiS itself also plays an important role in the CSE for several reasons. SiS, first detected by \citet{1975ApJ...199L..47M}, is one of the major silicon carriers and is believed to form in the dense and hot stellar atmosphere \citep{1973A&A....23..411T,1994ApJ...420..863B}. Indeed, this molecule is found to be the most abundant Si-bearing species in the CSE \citep{2012A&A...543A..48A}. In the dust formation zone (0\arcsec$\!\!$.1$<r<$0\arcsec$\!\!$.4) where the temperature is lower than the critical temperature of $\sim$1200~K \citep[e.g.,][]{2000ApJ...543..861M}, the SiS gas will condense to form dust grains due to its refractory character. Beyond this zone, SiS abundances are expected to decrease in the outer part of the CSE where SiS molecules are likely to be photodissociated. Hence, the quasi-thermal\footnote{In this work, the term ``quasi-thermal'' emission is used for lines not obviously showing maser amplification. This may occur not only in case of truly thermal emission but also in case of optically thin lines connecting states with inverted populations, which are not amplifying a strong radio continuum source in the background. The term also includes emission from non-inverted level populations, when spatial gas densities are too small to achieve thermalization.} emission of SiS can be used to investigate the molecular environment around IRC +10216. Previous interferometric SiS observations toward this CSE have yielded fruitful results. The $J$=5$\to$4 and $J$=6$\to$5 maps were found to display a centrally peaked morphology with a diameter of $\sim$18\arcsec\,\citep{1993AJ....105..576B,1995Ap&SS.224..293L}. The linear polarization of SiS (19$\to$18) was mapped with the Submillimeter Array (SMA), and has been interpreted in terms of a radial magnetic field configuration \citep{2012ApJ...751L..20G}. Combined Array for Research in Millimetre-wave Astronomy (CARMA) observations of the SiS $J$=14$\to$13 v=0 and v=1 lines were carried out by \citet{2014MNRAS.445.3289F}. They showed a very compact and nearly circular distribution with full width at half maximum (FWHM) sizes of less than 1\arcsec\,centered at the stellar position. The v=1, $J$=17$\to$16 and $J$=19$\to$18 lines of Si$^{34}$S and $^{29}$SiS are found to be confined to a region within 1\arcsec\,of the star and their profiles are very narrow, with expansion velocities up to $\sim$5~\kms\,\citep{2009ApJ...692.1205P}, suggesting that they arise from the dust formation region where the stellar wind is still being accelerated. Recent observations with the Atacama Large Millimeter/submillimeter Array (ALMA) resulted in the detection of the SiS $J$=15$\to$14 lines from highly vibrationally excited states up to v=7, and yielded very compact spatial distributions \citep{2015ApJ...805L..13V}, indicating that they are excited only in the innermost shells. Other highly excited SiS transitions (including isotopologues) have also been studied with ALMA, showing that they are concentrated at the stellar position with an FWHM size of less than 1\arcsec\,\citep{2015A&A...574A...5D}. These studies mainly focused on the very inner regions. However, the spatial distributions of the emission from the two lowest rotational SiS lines in IRC +10216's CSE are still poorly known. These two transitions have very low upper energy temperatures of $<$5~K above the ground state. With the Einstein coefficients for spontaneous emission and the collisional rate coefficients in the Leiden Atomic and Molecular database \citep[LAMBDA;][]{2005A&A...432..369S}, we obtain critical densities of 9.6$\times 10^{2}$~cm$^{-3}$ for SiS (1$\to$0) and 6.1$\times 10^{3}$~cm$^{-3}$ for SiS (2$\to$1) at an assumed kinetic temperature of 20 K. With such low critical densities and upper energy levels, the two SiS lines can be easily excited even in the outer region ($\sim 10^{3}$~cm$^{-3}$) of the CSE, making these lines good probes of the bulk distribution of SiS. Therefore, we have undertaken interferometric observations of SiS (1$\to$0) and SiS (2$\to$1) to explore the maser hypothesis, and to characterize the radial distribution of SiS.

With this twofold motivation, we present new observations of SiS toward the CSE of IRC +10216 that confirm the presence of an SiS (1$\to$0) maser. Complementary observations of less abundant SiS isotopologues are employed to constrain the excitation and optical depth of the SiS (1$\to$0) maser. We also report results from monitoring observations spanning more than ten years to study the maser emission's time variability. Interferometric observations of the two lowest rotational SiS lines are used to study the nature of the SiS (1$\to$0) maser and the distribution of SiS quasi-thermal emission in the CSE of IRC +10216.

\section{Observations}
\subsection{Observations of SiS and its isotopologues with the Effelsberg-100 m telescope}\label{sec.effobs}
SiS (1$\to$0), SiS (2$\to$1), Si$^{34}$S (1$\to$0), and $^{29}$SiS (1$\to$0) were observed in a position-switching mode with the 100-m telescope at Effelsberg/Germany\footnote{The 100-m telescope at Effelsberg is operated by the Max-Planck-Institut f{\"u}r Radioastronomie (MPIFR) on behalf of the Max-Planck-Gesellschaft (MPG).}, during 2001 August, 2002 June, 2004 May, 2007 April, 2008 September, 2012 January and April, and 2013 March. The beam size is about 50\arcsec\,at 18 GHz, and 25\arcsec\,at 36 GHz. The strong continuum source 3C~286 was used to calibrate the spectral line flux densities according to its standard flux density \citep[2.9 Jy at 18 GHz, and 1.6 Jy at 36 GHz;][]{1994A&A...284..331O}, and the calibration uncertainties are found to be within 10\%. The quasar PG 0851+202 (OJ+287) was used as the pointing source, and the pointing was found to be accurate to about 5\arcsec. For observations at different epochs, different backends were employed. During 2001--2004, an autocorrelator with eight segments was used as the back end. Each segment had a bandwidth of 20~MHz with 4096 channels for the high spectral resolution observations and 40~MHz with 512 channels for other observations. For the more recent observations from 2012 to 2013, a fast Fourier transform spectrometer (FFTS) was employed with a bandwidth of 2 GHz and 32768 channels. The observed transitions, the integration time, and corresponding spectral resolutions are displayed in Tables~\ref{Tab:obs1} and \ref{Tab:obs2}.

All spectral line data were analyzed with the CLASS/GILDAS\footnote{http://www.iram.fr/IRAMFR/GILDAS} software package. All spectra have been inspected, and those with system temperature higher than 100 K were discarded from further analysis. Spectra of a given line observed at the same epoch were then averaged with weights proportional to the inverse square of the system temperature. First-order baselines were subtracted from each spectrum. Throughout the paper, velocities are all given with respect to the local standard of rest (LSR).
\subsection{Interferometric observations}
We also observed SiS emission in the CSE of IRC +10216 with the Australia Telescope Compact Array (ATCA\footnote{The ATCA is part of the Australia Telescope National Facility which is funded by the Australian Government for operation as a National Facility managed by CSIRO.}) and the Karl G. Jansky Very Large Array (VLA\footnote{The VLA is a component of the National Radio Astronomy Observatory which is a facility of the National Science Foundation operated under cooperative agreement by Associated Universities, Inc.}). Table~\ref{Tab:interferometer} summarises these observations.
\subsubsection{ATCA}
SiS (1$\to$0) was observed with six ATCA 22-m antennas in the 1.5D, H214, and 6C array configurations on 2006 April 11, May 11, and June 26, respectively (project code C1560). The AT backend was used to record a total bandwidth of 4.1~MHz, with a channel width of 4 kHz, corresponding to 0.065~\kms\,at this frequency. PKS 1253$-$055 was used as the bandpass calibrator; PKS0912+029 was the gain and phase calibrator, and PKS 1934-638 was observed to calibrate the absolute flux density. The telescopes' pointing was checked every $\sim$2.5 hours toward the gain/phase calibrator. We estimate the absolute flux density to be accurate to about 10\%. The on-source time was about 6.5 hours in each array configuration, respectively. For our work, we combine the data observed with the three configurations to study the SiS (1$\to$0) emission. Three weightings, including natural weighting (robust=2), Briggs weighting (robust=0) and uniform weighting (robust=$-$2), were used in the imaging process. 

\subsubsection{VLA}\label{sec.jvla}
This work makes use of two different VLA projects to study the two low-$J$ SiS transitions. The SiS (1$\to$0) data were taken from project 11B-147 while the SiS (2$\to$1) data were obtained from the VLA's science verification data\footnote{see http://casa.nrao.edu/Data/EVLA/IRC10216/day2\_TDEM0003\_10s\_norx.tar.gz}. These data were observed with the VLA D-array configuration on 2011 December 8 and 2010 April 26, respectively.

The SiS (1$\to$0) observations were carried out with 26 VLA antennas, while the SiS (2$\to$1) observations are conducted with 19 VLA antennas. K- and Ka-band receivers were employed as front ends for the two transitions. The correlator setups provided a bandwidth of 8 GHz. SiS (1$\to$0) falls in a 64 MHz spectral window with 512 channels, while SiS (2$\to$1) falls in a 8 MHz spectral window with 64 channels. Channel widths are in both cases 125 kHz, which corresponds to 2.064~\kms\,at 18 GHz and 1.032~\kms\,at 36 GHz. During these observations, J1331+3030 (3C286) and J0954+1743 were observed as the absolute flux and gain calibrators, respectively, while J1229+0203 was used as a bandpass calibrator. These observations took approximately 3 hours for SiS (1$\to$0) and 2 hours 40 minutes for SiS (2$\to$1).

The data reduction was performed using the Common Astronomy Software Applications package \citep[CASA\footnote{http://casa.nrao.edu/}, version 4.7.0,][]{2007ASPC..376..127M}. Hanning smoothing was applied to the VLA SiS (1$\to$0) data to eliminate the Gibbs ringing effects, which led to a coarser spectral resolution by a factor of two. Briggs weighting (robust=0.5) was used to balance sensitivity and angular resolution in the clean processes. Using line-free channels, we also obtained spatially compact continuum emission centered on 36.35 GHz from the same archival data as SiS (2$\to$1), while uniform weighting was used in an attempt to achieve higher angular resolution.

\section{Results}
\subsection{Spectra observed with the Effelsberg-100 m telescope}\label{sec.effspec}
Figures~\ref{Fig:maser} and \ref{Fig:thermal} display the observed spectra of SiS (1$\to$0), SiS (2$\to$1), $^{29}$SiS (1$\to$0), and Si$^{34}$S (1$\to$0). As shown in Fig.~\ref{Fig:maser}a, we divide the SiS (1$\to$0) spectra into three velocity components, marked as ``blue'' ([$-$41,$-$38]~\kms), ``central'' ([$-$38,$-$15]~\kms), and ``red'' ([$-$15,$-$12]~\kms). For the 2001 Aug. 14 and 2004 May 2 epochs, the ``blue'' and ``red'' spikes are each fitted by assuming one single Gaussian component to get their velocities ($v_{\rm lsr}$), line widths ($\Delta v$), and peak flux densities ($S_{\nu}$). For other epochs, when the channel widths are too large to resolve the spike profiles, we use the peak emission channel to obtain the observed parameters of the ``blue'' and ``red'' components. For the ``central'' velocity component, we only obtain its average flux density. The SiS (2$\to$1), $^{29}$SiS (1$\to$0), and Si$^{34}$S (1$\to$0) transitions are fitted with the SHELL routine in CLASS to derive the line parameters including systemic velocity ($v_{\rm s}$), expansion velocity ($v_{\rm exp}$) and integrated flux density ($\int S_{\nu}{\rm d}v$). The results are listed in Tables~\ref{Tab:obs1} and \ref{Tab:obs2}. With the highest velocity resolution, the blue-shifted SiS (1$\to$0) spike is found to have a narrow line width of $\sim$0.4~\kms\, and shows a larger peak flux density ($\sim$1.09 Jy) than results previously reported \citep[less than 0.4 Jy; e.g.,][]{1983ApJ...267..184H,1985A&A...147..143H,2015A&A...574A..56G}. The systemic velocity (around $-$26.5~\kms) and expansion velocities (12.9$-$15.1~\kms) derived from SiS (2$\to$1), $^{29}$SiS (1$\to$0), and Si$^{34}$S (1$\to$0) are well consistent with previous studies \citep[e.g.,][]{2000A&AS..142..181C,2015A&A...574A..56G}.

As pointed out by \citet{1975ApJ...197..603M}, flat-topped and U-shaped profiles stem from optically thin spatially unresolved and resolved emission, respectively. The line profiles of $^{29}$SiS (1$\to$0) and Si$^{34}$S (1$\to$0) are flat-topped (see Fig.~\ref{Fig:thermal}), indicating that the two transitions are optically thin and that their emitting regions are not resolved by the 50\arcsec\,FWHM beam. In contrast to those two lines, the spectral profile of SiS (1$-$0) is distinct, showing two narrow spikes with the blue-shifted one being narrower and more prominent, which suggests that the two spikes are masers (their optical depths are negative, details are given in Sect.~\ref{dis.gain}). SiS (2$\to$1) shows a U-shaped profile, suggesting that it is optically thin and that the line emission is resolved with our 25\arcsec\,angular resolution. This is consistent with the extent of the SiS (2$\to$1) distribution, as discussed in Sect.~\ref{sec.thermal}.

Given the range of rotational temperatures found for other molecules in the CSE of IRC +10216 \citep[e.g.,][]{2015A&A...574A..56G}, i.e., 10~K to 40~K, we adopt these values for our analysis of SiS. Under local thermodynamic equilibrium (LTE) conditions, the beam-averaged ($\sim$25\arcsec) SiS column density derived from SiS (2$\to$1) is estimated to be $(8.5\pm 3.5)\times 10^{14}$~cm$^{-2}$. Since the ``central'' velocity component of SiS (1$\to$0) is likely quasi-thermal (see Sects.~\ref{sec.thermal} and \ref{dis.gain}), we can use its line temperature in combination with the line width (i.e. twice the expansion velocity of 14~\kms) to derive the beam-averaged ($\sim$50\arcsec) SiS column density as $(7.5\pm 4.5)\times 10^{14}$~cm$^{-2}$. Similarly, the beam-averaged ($\sim$50\arcsec) column densities of $^{29}$SiS and Si$^{34}$S are (4.6$\pm$2.6)$\times 10^{13}$~cm$^{-2}$ and $(5.2\pm3.0)\times10^{13}$~cm$^{-2}$, respectively, for the two adopted limiting rotation temperatures.

Assuming that SiS (1$\to$0), $^{29}$SiS (1$\to$0), and Si$^{34}$S (1$\to$0) are optically thin and have similar excitation temperature for the ``central'' velocity component, we can estimate the [$^{28}$Si/$^{29}$Si] and [$^{32}$S/$^{34}$S] isotopic ratios from their line ratios. Based on Fig.~\ref{Fig:thermal} and Table~\ref{Tab:obs2}, the flux densities of $^{29}$SiS (1$\to$0) and Si$^{34}$S (1$\to$0) are 5.5$\pm$0.9 mJy and 5.2$\pm$0.6 mJy, respectively. According to Table~\ref{Tab:obs1}, the average flux density of the ``central'' velocity component of SiS (1$\to$0) is found to be 88.3$\pm$11.5 mJy. Since these transitions of SiS and its isotopologues are close in frequency, the differential beam dilution effect on their ratios is negligible compared to other uncertainties. We then obtain the [$^{28}$Si/$^{29}$Si] and [$^{32}$S/$^{34}$S] isotopic ratios to be 16.1$\pm$3.4 and 17.0$\pm$3.0, respectively. These values agree with [$^{28}$Si/$^{29}$Si]$\sim$18 and [$^{32}$S/$^{34}$S]$\sim$20 derived by previous studies \citep{1988A&A...190..167K,2000A&A...357..669K,2000A&AS..142..181C,2008ApJS..177..275H,2012A&A...543A..48A}.

\subsection{Effelsberg-100 m monitoring observations of SiS (1$\to$0): time variability}
\begin{figure*}[!htbp]
\centering
\includegraphics[width = 0.8 \textwidth]{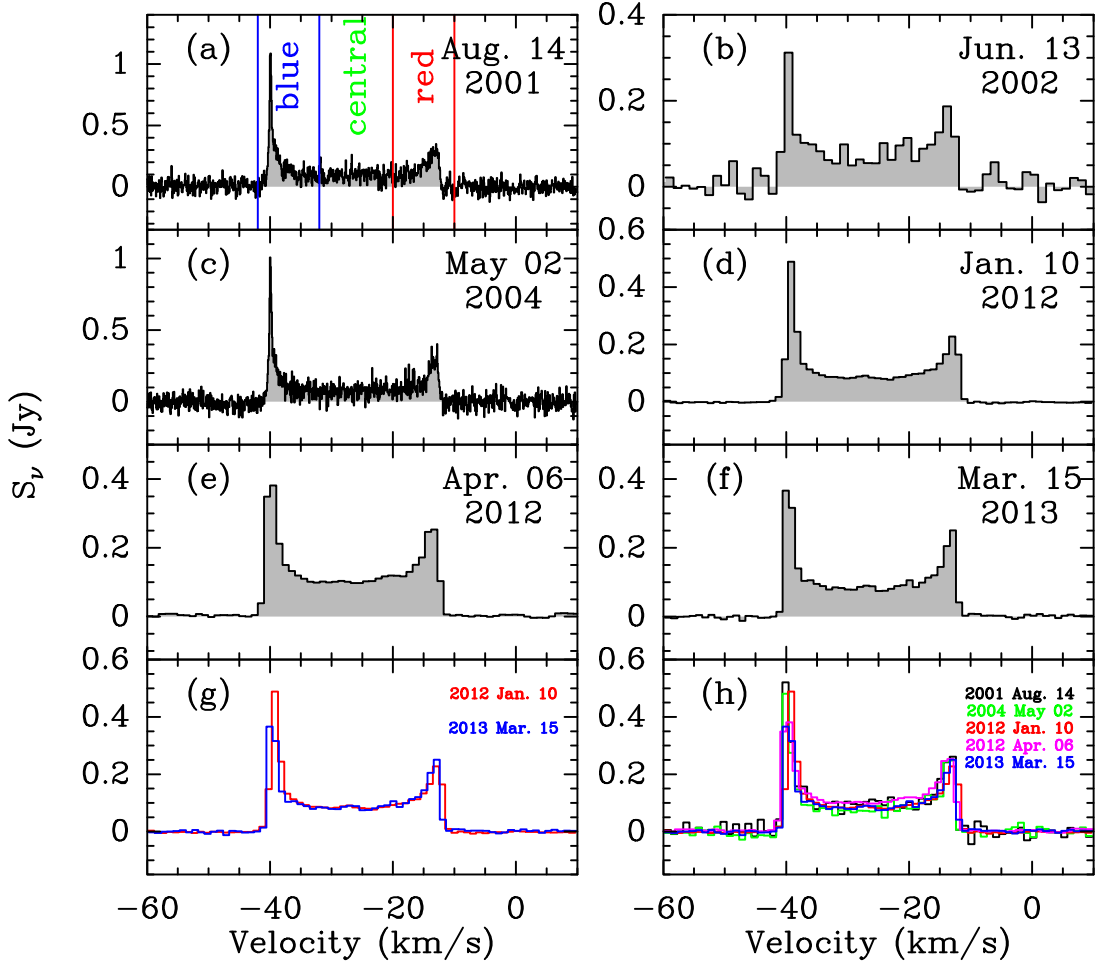}
\caption{{SiS (1$\to$0) spectra obtained with the Effelsberg-100 m telescope at the different epochs given in the upper right of each panel. In the lowest panels, the different colors correspond to the observing dates which are also indicated in the upper right of each panel. Figure~\ref{Fig:maser}g presents the two spectra with the largest difference in flux densities. In Fig.~\ref{Fig:maser}h, all spectra are overlaid after being smoothed to a channel width of 1.01\kms\, to provide an easily accessible measure of variations in line intensity between different epochs. The line center has been assumed to be non-variable.} \label{Fig:maser}}
\end{figure*}

\begin{figure*}[!htbp]
\centering
\includegraphics[width = 0.4 \textwidth]{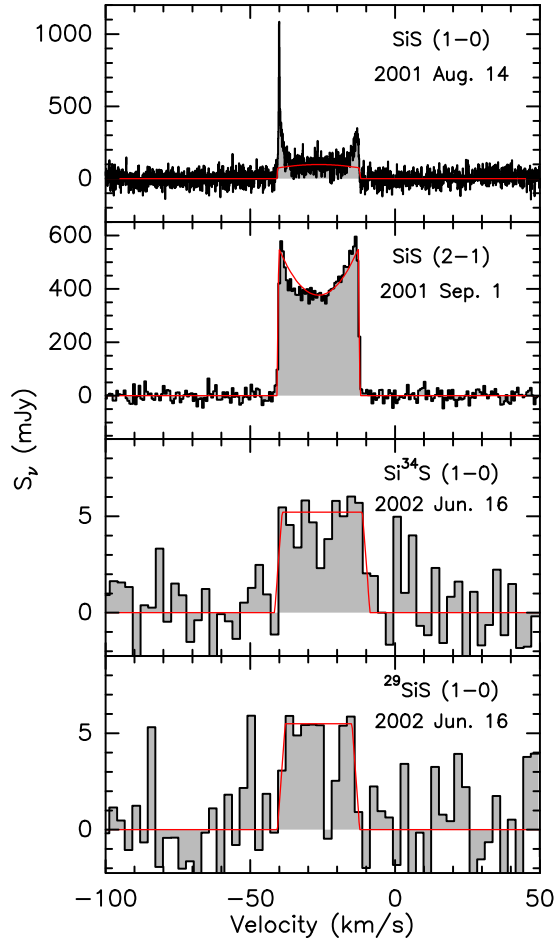}
\caption{{The observed Effelsberg-100 m spectra (black lines) of SiS (1$\to$0), SiS (2$\to$1), Si$^{34}$S (1$\to$0), and $^{29}$SiS (1$\to$0) with labels indicated in the upper right of each panel. Si$^{34}$S (1$\to$0), and $^{29}$SiS (1$\to$0) have been smoothed to have a channel width of $\sim$2.64~\kms . The red lines represent the fitted results through the ``SHELL'' routine in CLASS.} \label{Fig:thermal}}
\end{figure*}

\begin{figure*}[!htbp]
\centering
\includegraphics[width = 0.6 \textwidth]{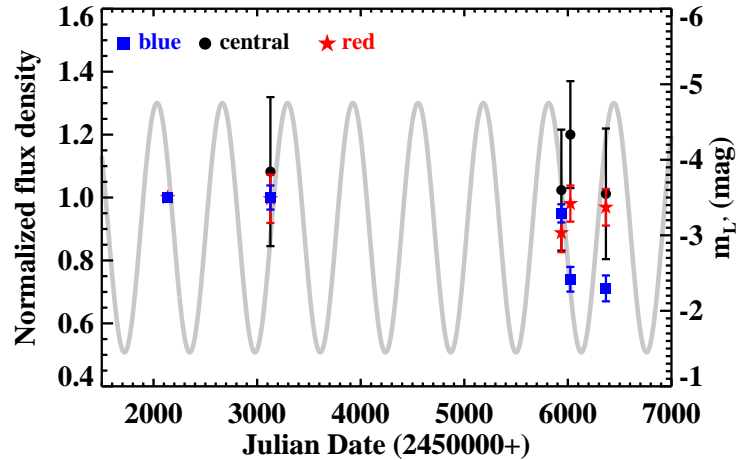}
\caption{{Normalized flux densities of the red-shifted spike (red pentagrams), the ``central'' component (black circles), and the blue-shifted spike (blue squares) of SiS (1$\to$0) in Fig.~\ref{Fig:maser}h to the measurement on 2001 Aug. 14 (JD 2452136) as a function of Julian date. The grey-shaded line represents the $L'$-band (3.76~$\mu$m) light curve of \citet{1992A&AS...94..377L} updated with a period of 630 days and the maximum brightness Julian date of 2454554 \citep{2012A&A...543A..73M}.} \label{Fig:var}}
\end{figure*}

Figure~\ref{Fig:maser} shows the spectra of SiS (1$\to$0) observed with the Effelsberg-100 m telescope at six different epochs from 2001 Aug. 14 (JD 2452136) to 2013 Mar. 15 (JD 2456367). According to Table~\ref{Tab:obs1}, the velocities of the blue-shifted and red-shifted spikes show little variation (by less than a channel width of 1.01\kms), demonstrating that these spikes have no significant velocity drift. It is known that the ``central'' velocity parts of the line profile are dominated by emission from the outer extended parts of the CSE. Thus, pointing errors may affect the flux density of the ``central'' component less than those of the blue-shifted and red-shifted spikes because they arise from a more compact region (less than 1\% of the Effelsberg beam, see Sect.~\ref{sec.maser}). Meanwhile, our typical pointing errors are found to be about 5\arcsec, $\sim$10\% of the FWHM beam size, and the quasar PG 0851+202 (OJ+287, the pointing source) is not extremely close to IRC +10216 (the angular distance is about 15\arcdeg). Fortunately, the blue-shifted and red-shifted spikes are not only compact, but also arise from almost the same line of sight directly toward the late-type star (see Sect.~\ref{sec.maser}). This implies that even pointing errors will not lead to any significant change in their {\it relative} intensity. Figure~\ref{Fig:maser}g only gives the two spectra with the largest difference in flux densities to visualize the amplitude of the time variability. In order to quantify the variation of the SiS (1$\to$0) spikes, we have resampled the spectra at five epochs into the same channel width of 1.01~\kms. They are presented in Fig.~\ref{Fig:maser}h. Figure~\ref{Fig:var} gives normalized flux densities of the two spikes and the ``central'' velocity component with respect to the epoch of 2001 Aug. 14. We note that the flux densities of the red-shifted spike and the ``central'' velocity component vary by less than 10\%\,and 20\%. The flux density difference of the blue-shifted component reaches a significant level of nearly 30\%, which supports the time variability of the blue-shifted SiS (1$\to$0) spike, proposed by previous observations with the NRAO 140-foot telescope at Green Bank and the Onsala-20 m telescope \citep{1984ApJ...286..276N,1985A&A...147..143H}. Although IRC +10216 is known to be a Mira variable with a period of $630\pm 3$ days at infrared wavelengths \citep{1992A&AS...94..377L,2012A&A...543A..73M} and a period of 535$\pm$50 days at cm continuum bands \citep{2006A&A...453..301M}, the number of epochs of observing that we now have is too small to allow for a search for periodic variations.

\subsection{Interferometer observations}\label{sec.interferometer}
\subsubsection{The VLA 36.35 GHz continuum}
\begin{figure*}[!htbp]
\centering
\includegraphics[width = 0.8 \textwidth]{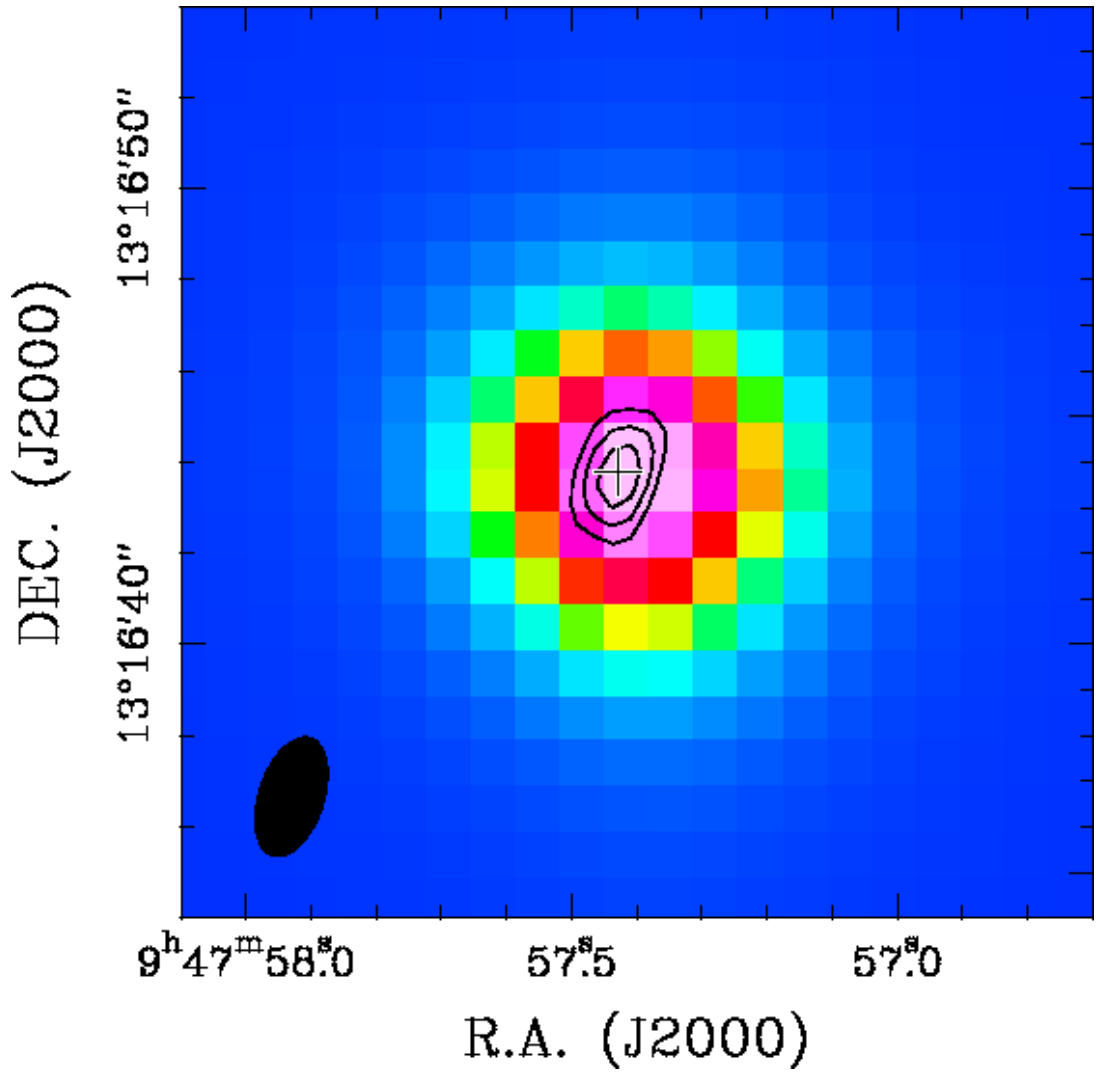}
\caption{{The VLA 36.35 GHz continuum emission contours overlaid on the 2MASS $J$ band image. The contours start at 4.0 mJy~beam$^{-1}$ (5$\sigma$) and increase by 1.8 mJy~beam$^{-1}$. The fitted position of the VLA 36.35 GHz continuum emission is marked by the cross. The synthesized beam is shown in the lower left.} \label{Fig:cont}}
\end{figure*}

Figure~\ref{Fig:cont} shows the 36.35 GHz continuum emission overlaid on the 2MASS $J$ band image, which shows that the 36.35 GHz continuum peak coincides with the peak of the infrared emission. The 36.35 GHz continuum emission stems from the radio photosphere of IRC +10216 \citep[e.g.,][]{2006A&A...453..301M,2012A&A...543A..73M} and is the only detected source in the map. Based on Gaussian fitting to the 36.25 GHz brightness distribution with the ``IMFIT'' task in CASA, we have determined its properties. The integrated flux density is found to be 11.0$\pm$1.3 mJy with a peak of 9.26$\pm$0.66 mJy~beam$^{-1}$. Note that calibration uncertainties are not included in the given errors. The integrated flux density at 36.35~GHz is consistent with an optically thick blackbody emission model \citep[see Fig.~2 of][]{2006A&A...453..301M}. The source is centered at $\alpha_{\rm J2000}$= 09$^{\rm h}$47$^{\rm m}$57$^{\rm s}\!\!$.428$\pm$0$^{\rm s}\!\!$.005 and $\delta_{\rm J2000}$=$+$13$^{\circ}$16\arcmin43\arcsec$\!\!$.79$\pm$0\arcsec$\!\!$.14. Taking a proper motion of (35, 12) mas yr$^{-1}$ in the eastward and northward directions into account \citep{2012A&A...543A..73M}, our measured position matches the positions of the 43 and 650 GHz continuum emission within uncertainties \citep{2012A&A...543A..73M,2015A&A...574A...5D}. The fitted convolved source size (2\arcsec$\!\!$.83$\pm$0\arcsec$\!\!$.25 $\times$1\arcsec$\!\!$.70$\pm$0\arcsec$\!\!$.10) is comparable to the synthesized beam (2\arcsec$\!\!$.74$\times$1\arcsec$\!\!$.48), indicating that the continuum source is unresolved as expected, since it has a diameter of about 83 milli-arcsec at 43 GHz \citep{2012A&A...543A..73M}.

\subsubsection{SiS (1$\to$0) maser action confirmed by ATCA}\label{sec.maser}
Figure~\ref{Fig:masermap} presents two channel maps of the SiS (1$\to$0) emission observed with ATCA. In Figs.~\ref{Fig:masermap}a---\ref{Fig:masermap}d, the emission in the northwest is likely affected by artifacts (see Appendix~\ref{a.psf}), because such an extended structure is not found in the VLA SiS (1$\to$0) map which has higher sensitivity and better uv coverage (see Sect.~\ref{sec.thermal}). Nevertheless, both the Briggs (robust=0) and uniform weighted images show a consistent distribution with compact emission around the star (see Figs.~\ref{Fig:masermap}c and \ref{Fig:masermap}e), and suggest that the emission is less affected by the artifacts. Figure~\ref{Fig:sis10maser} shows the spectra from the peak of the compact emission with different weightings applied. In this plot, two spikes stand out, which is similar to the single-dish SiS (1$\to$0) spectrum. In order to obtain the observed parameters, we use the ``IMFIT'' task in CASA to fit the compact emission which peaks at $-$39.862~\kms. Consequently, the peak flux density of the compact emission is found to be 923$\pm$23 mJy~beam$^{-1}$ in the uniform weighted image, which corresponds to a brightness temperature of 3850$\pm$100 K. The brightness temperature is still suffering from beam dilution effects, so the actual brightness temperature is, by all means, much higher than the rotational temperature ($\sim 2000$~K) of SiS in the innermost regions \citep[e.g.,][]{2012A&A...543A..48A,2015MNRAS.453..439F}, which unambiguously confirms the maser nature.

The maser emission is centered at $\alpha_{\rm J2000}$=09$^{\rm h}$47$^{\rm m}$57$^{\rm s}\!\!$.4249$\pm$0$^{\rm s}\!\!$.0001 and $\delta_{\rm J2000}$=$+$13$^{\circ}$16\arcmin43\arcsec$\!\!$.8949$\pm$0\arcsec$\!\!$.086. The quoted errors reflect the statistical uncertainties from Gaussian fitting. More realistically, the \textit{systematic} uncertainties due to the phase calibration process are difficult to quantify, but should be of the order of $0\rlap{.}''1$. Our position is offset by only ($\Delta \alpha, \Delta \delta$)=($0\rlap{.}''009$, $0\rlap{.}''079$) from the astrometric position of IRC +10216's photosphere resulting from higher resolution VLA observations \citep{2012A&A...543A..73M}, favoring that the blue-shifted component is directly located in front of the star. The velocity of the maser corresponds to an expansion velocity of 13.362$\pm$0.065~\kms\,by adopting a systemic velocity of $-$26.5~\kms\,\citep[e.g.,][]{2000A&AS..142..181C}. The expansion velocity indicates that the maser is likely to be produced in an {\it almost} fully accelerated shell. This suggests that the SiS (1$\to$0) maser is formed outside the innermost regions ($<$10R$_{\star}$) where the high-$J$ SiS maser candidates and submillimeter HCN masers are located \citep{2000ApJ...528L..37S,2003ApJ...583..446S,2006ApJ...646L.127F}. On the other hand, the peak brightness temperature of the red-shifted spike reaches 185$\pm$18 K in the Briggs (robust=0) weighted spectrum (see Fig.~\ref{Fig:sis10maser}). When beam dilution effects are taken into account, the actual brightness temperature may also exceed the rotational temperature ($\sim 600$~K) of SiS at $\sim$10R$_{\star}$ \citep{1994ApJ...420..863B,2012A&A...543A..48A,2015MNRAS.453..439F}, so the red-shifted spike may also exhibit maser action. However, we do not perform further analysis of the red-shifted spike due to contamination by artifacts caused by strong sidelobes and its relatively low signal-to-noise ratio.

\subsubsection{SiS quasi-thermal emission revealed by the VLA}\label{sec.thermal}
As mentioned in Sect.~\ref{sec.maser}, the blue-shifted and red-shifted spikes are likely affected by population inversion, so we only use the ``central'' velocity component to study the quasi-thermal SiS emission of the CSE in this section. In the following, we make use of the VLA data solely to study SiS quasi-thermal emission in the CSE of IRC +10216. It is worth noting that the observations presented here are interferometer-only data without the complementary short-spacing information in the UV plane. However, the shortest baseline of our VLA observations is about 35 m, corresponding to the largest reliable angular scale of about 98\arcsec\, at 18 GHz and 49\arcsec\,at 36 GHz, which indicates that the missing flux problem should not be serious for our detected SiS emission (lying within a radius of $\sim$11\arcsec, see results given below).

Figures~\ref{Fig:sismom}a and \ref{Fig:sismom}b display the integrated intensity maps of SiS (1$\to$0) and SiS (2$\to$1). Both integrated intensity maps show, to the first order, a circular and centrally peaked flux density distribution. The two lines peak at the same position which coincides with the 36.35 GHz continuum source. Figures~\ref{Fig:sismom}c and \ref{Fig:sismom}d give the SiS (1$\to$0) and SiS (2$\to$1) spectra from the peaks which are indicated by the crosses in Figs.~\ref{Fig:sismom}a and \ref{Fig:sismom}b. Both spectra show two spikes around $-$40~\kms\,and $-$13~\kms. The brightness temperatures of the spikes in Fig.~\ref{Fig:sismom}c are much lower than those in Fig.~\ref{Fig:sis10maser}, which is mainly due to the greater beam dilution in the larger VLA synthesized beams. Similar to SiS (1$\to$0), the blue-shifted spike (around $-$39.5~\kms) of SiS (2$\to$1) is stronger than the red-shifted spike (around $-$13~\kms) in Fig.~\ref{Fig:sismom}d. In view of a (to zeroth order) spherical envelope, potentially slightly opaque spikes (see Sect.~\ref{dis.gain} for the $J$=1$\to$0 line) should imply that the red-shifted spike would be more intense, because the self-absorption by comparatively low density and temperature foreground gas with low excitation temperatures should only affect the blue-shifted gas in front of the star. For optically thick transitions, we would then expect a higher intensity on the red-shifted spikes, opposite to what is observed (Fig.~\ref{Fig:sismom}d). Whether amplification of the stellar continuum (see Sect.~\ref{dis.gain}) is playing a role as for the $J$=1$\to$0 line or whether deviations from spherical geometry also Sect.~\ref{dis.gain} cause this discrepancy remains open. We also note that the flux densities of the two SiS (2$\to$1) spikes are nearly identical in our Effelsberg-100 m measurements (see Fig.~\ref{Fig:thermal}), which are dominated by gas located directly in front of and behind the star. Toward the ``central'' velocity component, we find that the brightness temperatures of the two transitions are lower than 5~K. Here, we also make use of the non-LTE code, RADEX \citep{2007A&A...468..627V}, to evaluate the optical depth of the two transitions. At radii of about 3\arcsec, which corresponds to the beam sizes (see Table~\ref{Tab:interferometer}), \citet{2012A&A...543A..48A} and \citet{1988ApJ...326..832K} find a gas temperature of $T_{\rm kin} \sim$120 K and an H$_{2}$ number density of $n({\rm H}_{2})\sim 10^{5}$~cm$^{-3}$. By assuming the column density per line width to be $N/\Delta v$=3.6$\times 10^{13}$~cm$^{-2}$~(\kms)$^{-1}$ (based on the results in Sect.~\ref{sec.effspec}), the RADEX calculations confirm that the optical depths of the two lines are both lower than 0.1. Even when we use a much higher column density per line width to be $N/\Delta v$=1.0$\times 10^{15}$~cm$^{-2}$~(\kms)$^{-1}$, the optical depths are still lower than 0.1. Therefore, we conclude that the ``central'' velocity component is optically thin in SiS (1$\to$0) and SiS (2$\to$1). 

The azimuthally-averaged, velocity-integrated radial intensity profiles of SiS (1$\to$0) and SiS (2$\to$1) are investigated and shown in Fig.~\ref{Fig:intprofile}. We find that these profiles can be fit with a two-component Gaussian, indicative of both a compact and an extended component. From SiS (1$\to$0), the compact one is found to have an FWHM of 3\arcsec$\!\!$.8$\pm$2\arcsec$\!\!$.3, while the extended one has an FWHM of 12\arcsec$\!\!$.5$\pm$2\arcsec$\!\!$.5. For SiS (2$\to$1), we obtain an FWHM of 2\arcsec$\!\!$.3$\pm$2\arcsec$\!\!$.2 for the compact one and 12\arcsec$\!\!$.6$\pm$0\arcsec$\!\!$.9 for the extended one. The compact component is therefore not resolved by our observations, and the FWHM difference of the compact component derived from SiS (1$\to$0) and SiS (2$\to$1) is mainly due to the different synthesized beams of the images. On the other hand, the small size of the compact component implies that it likely corresponds to SiS emission arising from the innermost acceleration shells which are regarded as the dust formation zone of IRC +10216 \citep[e.g.,][]{2014MNRAS.445.3289F}. The presence of such a compact component agrees with the fact that SiS is a parent molecule that is formed close to the star through thermodynamical equilibrium chemistry \citep{1973A&A....23..411T,1975ApJ...199L..47M}. The FWHM of the extended component is nearly identical for the two transitions. Deconvolved widths are found to be 12\arcsec$\!\!$.3 and 12\arcsec$\!\!$.0. The radial distributions of the two transitions are more extended than those of high $J$ SiS transitions (see Sect.~\ref{sec:intro}), which is attributable to the lower critical densities and lower energies of the rotational levels involved in these lines. Figure~\ref{Fig:intprofile} shows that SiS is detected beyond a radius of 11\arcsec, suggesting that SiS can trace the history of mass loss for at least the past $\sim$470 yr. Nevertheless, SiS emission is not as extended as CO which is detected out to a radius of $\sim$180\arcsec\,\citep{2003ApJ...582L..39F,2006ApJ...652.1626F,2015A&A...575A..91C}. This is likely because the interstellar radiation field dissociates SiS more effectively than CO due to the difference in column densities and thus self-shielding \citep[e.g.,][]{1983ApJ...264..546M}, as well as the difference in ground state dissociation energies \citep[ground state dissociation energies of CO and SiS are 11.09 eV and 6.4 eV, table 2.1 of][]{1984inch.book.....D}. 

%%The spikes in Fig.~\ref{Fig:sismom}c have very narrow line widths (less than 1~\kms), indicating that they may also be affected by maser actions. However, their brightness temperatures are too low (20$-$30~K) to confirm the maser nature. Alternatively, there is a possible more extended SiS (2$-$1) emission than the 3\arcsec$\!\!$.33$\times$2\arcsec$\!\!$.01 beam. Then we can automatically see very pronounced blue- and red-shifted spikes toward the center, because most of the emission near the systemic velocity originates from outside the beam.

Figures \ref{Fig:sis10-ch} and \ref{Fig:sis21-ch} present the channel maps of SiS (1$\to$0) and SiS (2$\to$1). The emitting regions become larger as the velocities trend from the extreme velocities (around $-$40~\kms\,and $-$12~\kms) to the systemic velocity \citep[around $-26.5$~\kms,][]{2000A&AS..142..181C}, consistent with a spherical envelope having a terminal expansion velocity of $\sim$14~\kms. Comparing the channel maps of the SiS (1$\to$0) and SiS (2$\to$1) emission, one notes that emission from the former line covers a somewhat larger velocity range than that from the last, which is due to the coarser spectral resolution of SiS (1$\to$0) (see Sect.~\ref{sec.jvla}). Furthermore, we find that there are more asymmetric features in Fig.~\ref{Fig:sis21-ch} than in Fig.~\ref{Fig:sis10-ch}. This is because both angular resolution and spectral resolution are higher in the SiS (2$\to$1) data. \citet{1995Ap&SS.224..293L} found an elongation along an axis with PA of $\sim$20\arcdeg\,in their SiS (5$\to$4) map averaged over the velocity range from $-$28.5 to $-$21.5~\kms, but such a structure is not found in our maps. This is probably because the elongated structure stands out best in the highly excited lines. For the two lowest-$J$ lines, emission from such a structure is likely blended with emission from the ambient, low-excitation gas component, making it imperceptible in our data.

In the $-$30 to $-$21.8~\kms\,velocity range (see Fig.~\ref{Fig:sis21-ch}), SiS emission shows a complex morphology with several possible shell-like structures evident in individual $\approx 2$~\kms\,wide channel maps. At a radius of $\sim$12\arcsec, Figure~\ref{Fig:arc} shows an incomplete shell-like structure in the north-east of the CSE. The ratios between the maximum flux of the shell to the minimum flux in the inter-shell are 3.7, 3.0, 2.1, and 2.2 in the four panels. Furthermore, this incomplete shell also displays a red-shifted component in the velocity fields of both SiS (1$\to$0) and SiS (2$\to$1) emission (see Appendix~\ref{sec.mom}. The molecular shells are superposed on the smoothed Gaussian components in Fig.~\ref{Fig:sis21-ch} but hardly seen due to their low shell-intershell contrasts. In order to increase the contrasts and better visualize the molecular shells, double circular Gaussian components (corresponding to a compact and an extended component) are fitted to the observed emission channel by channel in Fig.~\ref{Fig:sis21-ch} and then subtracted, producing the results shown in Figure~\ref{Fig:sis21-resi}. The fitted FWHMs of the compact component are less than 4\arcsec, dominated by the synthesized beam. The extended component has different FWHMs in different channels, and the FWHMs vary from 5\arcsec$\!\!$.6 to around 15\arcsec. Intriguingly, incomplete detached molecular shells with signal-to-noise ratios of $>$3 are seen in both the continuum and two-component spectral-background-subtracted channel maps (see Fig.~\ref{Fig:sis21-resi}). Despite low signal-to-noise ratios, these structures extend continuously through several channels. The shells are asymmetric with strong emission at the north-east, and are non-concentric, which is consistent with Fig.~\ref{Fig:arc}. Furthermore, there are HC$_{3}$N shells detected at the same region \citep[see Fig.~1 of][]{2017A&A...601A...4A}. These facts indicate the existence of SiS shells. Although molecular shells in this CSE have been reported by many previous studies \citep[e.g.,][]{1993A&A...280L..19G,2000A&A...359..707M,2003ApJ...582L..39F,2006PNAS..10312274Z,2008ApJ...678..303D,2015A&A...574A...5D}, this is the first indication that such shells are detected in SiS emission. Moreover, there are several clumpy structures with signal-to-noise ratios of $>$5 in these detached shells, but these clumpy structures are not resolved by our synthesized beam. Similar clumpy structures have been also found by previous studies \citep[e.g.,][]{1995Ap&SS.224..293L,2003ApJ...582L..39F,2008ApJ...678..303D}.

Figure~\ref{Fig:sis21-pv} gives the SiS (2$\to$1) position-velocity (PV) diagrams for PA=35\arcdeg\,and PA=124\arcdeg. Overall, the PV diagrams display oval-shaped structures, similar to PV diagrams of other molecules \citep[e.g.,][]{2006ApJ...652.1626F,2015A&A...574A...5D}. Also, there is at least one arc feature in the PV diagrams, supporting the presence of expanding shells. Similar curved structures in PV diagrams are also found in the $^{13}$CO (6$\to$5) ALMA data \citep{2015A&A...574A...5D}. Around 0\arcsec\,offset, the brightest emission is found to be near $-$40~\kms\,and $-$13~\kms, because the optical depth reaches a maximum, while the velocity gradient becomes minimal once the terminal expansion velocity has been reached.

\section{Discussion}
A follow-up systematic search for potential SiS (1$\to$0) masers in other evolved stars was carried out with the Effelsberg-100 m telescope (Henkel, C., priv. comm.). During these observations, SiS was only detected in CRL 3068, CIT 6, and CRL 2688, but in none of these objects does the $J=1\to 0$ transition show narrow spikes similar to those observed in IRC +10216. Thus, the SiS (1$\to$0) maser in IRC +10216 is so far unique. We will discuss this maser in the following.
\subsection{The gain of the SiS (1$\to$0) maser}\label{dis.gain}
Here, we will discuss the gain of the SiS (1$\to$0) maser in two different ways. Firstly, taking the continuum emission from the star into account, the radiative transfer equation under the Rayleigh-Jeans limit (h$\nu$/k$\sim$1 K $\ll T_{\rm ex}$; h: Planck constant, k: Boltzmann constant) becomes:
\begin{equation}\label{f.rt}
T_{\rm line}=f(T_{\rm ex}-T_{\rm mol}-T_{\rm bg}-T_{\rm c})(1-e^{-\tau})\;,
\end{equation}
where $T_{\rm line}$ is the observed line brightness temperature, $T_{\rm ex}$ is the excitation temperature, $T_{\rm mol}$ is the brightness temperature of SiS line emission lying behind the maser, $T_{\rm bg}$ is the microwave background radiation which is equal to 2.7255$\pm$0.0006 K \citep{2009ApJ...707..916F}, $T_{\rm c}$ is the brightness temperature of the continuum emission from the star or the near stellar environment, $f$ is the filling factor $f=\Omega_{\rm s}/(\Omega_{\rm s}+\Omega_{\rm beam})$, $\Omega_{\rm s}$ the solid angle subtended by the source, and $\Omega_{\rm beam}$ the solid angle occupied by the telescope beam. In formula~(\ref{f.rt}), $T_{\rm c}$ only contributes to the sources in front of the star along the line of sight, while $T_{\rm c}$ is set to zero for other directions. Consequently, the blue-shifted component amplifies the stellar continuum, while the red-shifted component only amplifies the SiS (1$\to$0) emission arising from outer shells of the CSE and the microwave background radiation, which is much weaker. This leads to a brightness temperature of the blue-shifted spike that is much higher than for the red-shifted spike.

The size of the stellar radio emission was measured to be 83 milli-arcsec (diameter) according to VLA observations at 7 mm wavelength \citep[43 GHz;][]{2012A&A...543A..73M}. Here, we simply assume the radio emission at 18~GHz emanates from a source of the same size as at 7 mm wavelength. The radio continuum emission is known to be optically thick, so the brightness temperature is the same at all radio wavelengths, i.e., 1630 K \citep{2012A&A...543A..73M}. Assuming that $T_{\rm c}$ is much larger than $T_{\rm mol}$, $T_{\rm bg}$, and the absolute value of $T_{\rm ex}$, we can neglect $T_{\rm mol}$, $T_{\rm ex}$, and $T_{\rm bg}$ in formula~(\ref{f.rt}) to analyze the blue-shifted component of the maser that lies directly in front of the star. As already mentioned, the ATCA observations yield $T_{\rm line}$=3850$\pm$100 K for the blue-shifted spike under the synthesized beam size of 2\arcsec$\!\!$.62$\times$0\arcsec$\!\!$.34. Since the blue-shifted spike is beaming and amplifying the continuum emission, we assume its size to be the same as that of the stellar radio emission, which results in a beam filling factor of 1/130 for the ATCA synthesized beam size. This gives an optical depth of $-$5.7$\pm$0.1, which agrees with the previous prediction (based on much lower angular resolution single-dish data) of $-$5.4 by \citet{1983ApJ...267..184H}. We also note that neglecting $T_{\rm mol}$ may lead to a lower absolute value of the optical depth. However, the difference is relatively small. For instance, if we take $T_{\rm mol}$ as high as $T_{\rm c}$, i.e., 1630 K, the optical depth becomes $-$5.0.

Secondly, assuming that the transitions of SiS and its isotopologues have nearly the same excitation temperature, we can use line ratios at the extreme velocities to estimate the optical depth of these lines according to the known isotopic ratios [$^{28}$Si/$^{29}$Si] and [$^{32}$S/$^{34}$S] with the formula:
\begin{equation}\label{f.iso}
\frac{T_{\rm iso}}{T_{\rm main}}=\frac{1-e^{-\tau_{\rm m}/r}}{1-e^{-\tau_{\rm m}}}\;,
\end{equation}
where $T_{\rm iso}$ is the brightness temperature of $^{29}$SiS (1$\to$0) or Si$^{34}$S (1$\to$0), and $T_{\rm main}$ is the brightness temperature of the main isotope's line. $\tau_{\rm m}$ is the optical depth of SiS (1$\to$0), and $r$ is the isotopic ratio [$^{28}$Si/$^{29}$Si] or [$^{32}$S/$^{34}$S]. We also note that the assumption is only valid if the radiation field does not affect the excitation temperature. In a maser, the assumption of similar excitation temperatures can only hold if the maser is unsaturated (see below). As already discussed in Sect.~\ref{sec.effspec}, we use [$^{28}$Si/$^{29}$Si]=18 and [$^{32}$S/$^{34}$S]=20 for the following calculations. Toward the blue-shifted and red-shifted spikes, we adopt 1.09$\pm$0.05 Jy and 0.32$\pm$0.05 Jy as their peak flux densities, respectively, based on our single-dish observations with high velocity resolution on 2001 Aug. 14 and 2004 May 2 (see Table~\ref{Tab:obs1}). As a result, the line ratios $\frac{\rm SiS\;(1\to 0)}{\rm ^{29}SiS\;(1\to 0)}$ and $\frac{\rm SiS\;(1\to 0)}{\rm Si^{34}S\;(1\to 0)}$ are found to be 200$\pm$30 and 210$\pm$30 for the blue-shifted component, which gives optical depths of $-$3.9$\pm$0.2 and $-$3.8$\pm$0.2 according to formula (\ref{f.iso}). If an optical depth of $-$3.9 is adopted in formula~(\ref{f.rt}), this yields a size of 0\arcsec$\!\!$.27 for the blue-shifted component, which agrees with the fact that the maser is not resolved by our ATCA observations. Similarly, the line ratios $\frac{\rm SiS\;(1\to 0)}{\rm ^{29}SiS\;(1\to 0)}$ and $\frac{\rm SiS\;(1\to 0)}{\rm Si^{34}S\;(1\to 0)}$ of the red-shifted component are 58$\pm$13 and 62$\pm$12, leading to optical depths of $-$2.1$\pm$0.4 and $-$2.0$\pm$0.3. This indicates that the red-shifted component may also exhibit maser action, but its confirmation still needs further observations. In addition, the different optical depths between the blue-shifted and red-shifted components of the profile may be due to anisotropies in this CSE.  

With the two independent methods, all derived absolute values of optical depths are larger than unity but not very much larger. Therefore, we suggest that the SiS (1$\to$0) maser is unsaturated in the CSE of IRC +10216.

\subsection{The pumping mechanism of the SiS (1$\to$0) maser in IRC +10216}
Collisional and infrared pumping are two mechanisms to excite masers. Based on the density profile of \citet{1988ApJ...326..832K}, the SiS (1$\to$0) maser is likely formed in a region with an H$_{2}$ density of $>$10$^{4}$ cm$^{-3}$, indicating that collisions are important. However, the level populations of SiS producing the $J=1\to 0$ transition will become thermalized if collisional pumping is dominant, because the region where the maser is formed has a density much higher than the critical density ($\sim$9.6$\times 10^{2}$~cm$^{-3}$) of SiS (1$\to$0). Since the level populations involved in the maser cannot be, by definition, thermalized, we therefore propose infrared pumping as the main pumping mechanism.  

A pumping mechanism based on a process comprising ro-vibrational transitions has been introduced to explain the effect of infrared radiation on rotational excitation of diatomic molecules \citep{1975ApJ...197..603M,1977ApJ...218..687M,1980ApJ...236..823M,1981ApJ...245..891C}. This mechanism can be used to explain the SiS (1$\to$0) maser. Figure~\ref{Fig:energy} illustrates the most important transitions for producing a population inversion between the v=0, $J$=0 and v=0, $J$=1 levels of SiS. The infrared excitation in all of the ro-vibrational lines followed by decay back down to the ground vibrational state tends to lead to $\Delta J$=+2 steps in the rotational ladder\footnote{For SiS or other diatomic molecules with permanent dipole moment, the selection rule for electric dipole allowed ro-vibrational transitions is $\Delta$v$=\pm1,\;(\pm2,\;etc.)$, $\Delta J=\pm1$.}. If the infrared radiation field is dilute, the subsequent rotational transitions in the v=0 state lead to a general cascade back down to lower $J$ levels. The drainage slows down with lower $J$ because of the strong dependence of the Einstein A coefficient on $J$ ($A\sim J^{3}$). Hence, SiS molecules tend to cascade down to the v=0, $J$=1 level and pile up there, because of the very slow $J=1\to 0$ transition, compared to higher rotational transitions. In addition, there is a deficit in the v=0, $J$=0 state because it is depopulated by infrared excitation to the v=1, $J$=1 state followed predominantly by decay down to the v=0, $J$=2 state, so the pumping mechanism creates a population inversion between the v=0, $J$=0 and $J=1$ levels. In order to maintain this population inversion, the rate of infrared excitations out of the v=0, $J$=1 up to the v=1, $J$=2 state should be lower than the spontaneous decay rate in the v=0, $J$=2$\to$1 line. The proposed pumping mechanism therefore breaks down in the innermost shells where the intensity of the v=0, $J$=1 to v=1, $J$=2 line becomes large enough to violate this condition or where the density becomes high enough that collisions thermalize the rotational levels. This is consistent with our results that the SiS (1$\to$0) maser arises from an almost fully accelerated shell (see Sect.~\ref{sec.maser}).

We also note the absence of other low $J$ SiS masers in the ground vibrational state. For the SiS (1$\to$0) transition, the v=0, $J$=1 state has only one outlet to the v=1, $J$=2 state via infrared excitation, because the transition to the v=1, $J$=0 state can only fall immediately back to the v=0, $J$=1 state, which does not have any effect. The upper energy levels of other rotational lines have two outlets via infrared radiation, so it is therefore harder to maintain an inversion in those states. For example, the v=0, $J$=2 state can be excited to the v=1, $J$=1 or 3 state, and both of those states have alternative decay paths that do not simply fall back to the v=0, $J$=2 state.

Another potential mechanism for pumping SiS masers is based on overlaps between the ro-vibrational lines of SiS with those of other molecules. \citet{2006ApJ...646L.127F} proposed this mechanism to explain the $J$=11$\to$10, $J$=14$\to$13, and $J$=15$\to$14 SiS maser candidates that they detected in IRC +10216. In general, this process is likely to gives rise to line asymmetries in the rotational lines because infrared lines from other species would likely be slightly displaced in frequency from the SiS transitions. However, the SiS (1$\to$0) line shows two sharp peaks which have nearly the same expansion velocity. This indicates that overlapping infrared lines may not play a dominant role in pumping the SiS (1$\to$0) maser. We therefore favor infrared continuum emission, rather than infrared line overlaps, as the main pumping source for the SiS (1$\to$0) maser. 

CO (1$\to$0) has an Einstein A coefficient ($7.20\times 10^{-8}$~s$^{-1}$) similar to that ($6.99\times 10^{-8}$~s$^{-1}$) of SiS (1$\to$0), so our proposed mechanism may also predict CO (1$\to$0) masers. However, CO (1$\to$0) has not been found to show maser action. The infrared emission in outer shells, if not intense enough for efficient pumping of SiS, should also not be sufficient to pump a CO maser, neglecting any saturation effects. As pointed out by \citet{1980ApJ...236..823M}, the presence of CO (1$\to$0) masers depends upon the optical depths of the ro-vibrational lines (around 4.6 $\mu$m) not being very large. While in the inner region, CO ro-vibrational lines are optically thick, which will quench the maser.

\section{Summary and conclusion}
We have studied the maser and quasi-thermal emission of the v=0, SiS $J$=1$\to$0 and 2$\to$1 lines in the circumstellar envelope (CSE) of IRC +10216 by means of Effelsberg-100 m, ATCA, and VLA observations. This has led to the following main results:
\begin{itemize}
\item[1.] Based on the ATCA data, we find that the blue-shifted component ($v_{\rm lsr}$=$-$39.862$\pm$0.065~\kms) of the SiS ($1\to 0$) transition reaches a very high lower brightness temperature limit of 3850$\pm$100 K in the inner part of the CSE, unambiguously confirming the presence of a so far unique SiS (1$\to$0) maser in IRC +10216 more than thirty years after its first tentative assignment. The blue-shifted component, unresolved by our observations, is found to lie directly in front of the star. Its expansion velocity indicates that the maser is likely formed at an almost fully accelerated shell.

\item[2.] Our Effelsberg-100 m monitoring observations support variability in the blue-shifted component of the SiS (1$\to$0) maser. It is not yet clear whether this variability is ascribable to the strong variability of the central infrared source or to density inhomogeneities in the outflowing wind.

\item[3.] The quasi-thermal emission of the lowest-$J$ SiS transitions traces a much more extended emitting region than ever seen in previous high-$J$ SiS observations. Their distributions show that the SiS quasi-thermal emission consists of two components: one is very compact with a size of $<$3\arcsec, and the other extends out to an angular distance of $>$11\arcsec. The extended SiS emission shows that an incomplete shell-like structure is found in the north-east, which is indicative of existing SiS shells. Also, the extended SiS emission reveals a number of clumpy structures in this CSE. 

\item[4.] The gain of the SiS (1$\to$0) maser is estimated with two methods, one assuming amplification of the background continuum and the other using rare isotopologues in combination with known isotopic ratios. The derived absolute values of optical depths are larger than unity but not very much larger (about $-$5 for the blue-shifted and, if also inverted, about $-$2 for the red-shifted component), which suggests that the maser is unsaturated. The difference in opacities may hint at an asymmetry in the shell with respect to gas in front of and behind the central stellar object. The SiS (1$\to$0) maser can be explained in terms of ro-vibrational excitation caused by infrared pumping, and we propose that infrared continuum emission is the main pumping source.  
\end{itemize}

\acknowledgments
We appreciate the assistance of the Effelsberg 100-m, ATCA, and VLA operators during the observations. We gratefully acknowledge the anonymous referee for the insightful comments on the draft. Y. Gong acknowledges support by the MPG-CAS Joint Doctoral Promotion Program (DPP), the National Natural Science Foundation of China (NSFC) (grants nos. 11127903, 11233007, and 10973040), and the Strategic Priority Research Program of the Chinese Academy of Sciences (grant no. XDB09000000). This research made use of NASA's Astrophysics Data System.


%% for the paper.  Each keyword is check against the master list during
%% copy editing.  Individual instruments can be provided in parentheses,
%% after the keyword, but they are not verified.
\facilities{Effelsberg-100 m, ATCA, VLA} 
\software{GILDAS, CASA\citep{2007ASPC..376..127M}}

%{\it Facilities:} \facility{Effelsberg-100 m}, \facility{ATCA}, \facility{VLA}.



\clearpage
\floattable
\begin{deluxetable}{ccccccccccccc}
\tablecolumns{13}
\rotate
\tabletypesize{\scriptsize}
\centering
\tablewidth{0pc}
\tablecaption{Summary of observational parameters of SiS (1--0) at the Effelsberg-100 m telescope.\label{Tab:obs1}}
\tablehead{
          &           &           &               &        &        & \multicolumn{3}{c}{blue}    &central & \multicolumn{3}{c}{red}     \\
\cline{7-9} \cline{11-13}
\colhead{Frequency}&\colhead{Transition}&\colhead{$E_{\rm u}/k$}&\colhead{Epoch}      &\colhead{Ch.width}&\colhead{int.time}&\colhead{$v$}&\colhead{$\Delta v$}&\colhead{$S_{\nu}$} &\colhead{$S_{\nu}$}&\colhead{$\upsilon$}&\colhead{$\Delta v$}&\colhead{$S_{\nu}$}  \\
\colhead{(MHz)}      &   \colhead{}        &\colhead{(K)}        &  \colhead{}             & \colhead{(\kms)} & \colhead{(min)}  &  \colhead{(\kms)} & \colhead{(\kms)} & \colhead{(Jy)}      &    \colhead{(Jy)}      &  \colhead{(\kms)} & \colhead{(\kms)} & \colhead{(Jy)}  \\          
  \colhead{(1)}      &   \colhead{(2)}     & \colhead{(3)}        &  \colhead{(4)}          & \colhead{(5)}    & \colhead{(6)}    &  \colhead{(7)}    & \colhead{(8)}    & \colhead{(9)}       &    \colhead{(10)}      &  \colhead{(11)}   & \colhead{(12)}   & \colhead{(13)}}
\startdata
18154.9    &SiS (1$\to$0) &  1     & 2001 Aug. 14  & 0.08   & 26     &$-$39.9$\pm$0.1 & 0.4$\pm$0.1 & 1.094$\pm$0.046 & 0.090$\pm$0.050 & $-$13.3$\pm$0.1 & 1.6$\pm$0.1 & 0.323$\pm$0.046    \\
           &              &        & 2001 Aug. 14  & 1.01$^{*}$   & 26     &$-$39.9$\pm$1.0 & \nodata     & 0.516$\pm$0.014 & 0.095$\pm$0.016 & $-$13.3$\pm$1.0 & \nodata     & 0.259$\pm$0.014    \\
           &           &           & 2002 Jun. 13  & 1.29   & 29     &$-$39.7$\pm$1.3 & \nodata     & 0.312$\pm$0.021 & 0.068$\pm$0.023 & $-$13.8$\pm$1.3 & \nodata     & 0.187$\pm$0.021    \\
           &           &           & 2004 May   2  & 0.08   & 27     &$-$39.9$\pm$0.1 & 0.4$\pm$0.1 & 1.093$\pm$0.044 & 0.077$\pm$0.046 & $-$13.3$\pm$0.1 & 1.6$\pm$0.1 & 0.318$\pm$0.044   \\
           &              &        & 2004 May   2  & 1.01$^{*}$   & 27     &$-$39.9$\pm$1.0 & \nodata     & 0.516$\pm$0.014 & 0.092$\pm$0.016 & $-$13.3$\pm$1.0 & \nodata     & 0.258$\pm$0.014    \\
           &           &           & 2012 Jan. 10  & 1.01   & 343    &$-$39.8$\pm$1.0 & \nodata     & 0.490$\pm$0.002 & 0.087$\pm$0.005 & $-$13.1$\pm$1.0 & \nodata    & 0.230$\pm$0.002   \\
           &           &           & 2012 Apr.  6  & 1.01   & 122    &$-$39.5$\pm$1.0 & \nodata     & 0.382$\pm$0.004 & 0.102$\pm$0.008 & $-$13.2$\pm$1.0 & \nodata    & 0.254$\pm$0.004   \\
           &           &           & 2013 Mar. 15  & 1.01   & 127    &$-$39.3$\pm$1.0 & \nodata     & 0.367$\pm$0.004 & 0.086$\pm$0.008 & $-$13.3$\pm$1.0 & \nodata    & 0.251$\pm$0.004   \\ 
\enddata
\tablecomments{(1) the rest frequency of the corresponding transition; (2) the transition; (3) the upper energy temperature of the transition; (4) the epoch of the observations; (5) the channel width; (6) the on-source integration time of the observation; (7) the velocity of the blue-shifted spike; (8) the FWHM line width of the blue-shifted spike; (9) the flux density of the blue-shifted spike; (10) the average flux density of the ``central'' component within the velocity range from $-38$ to $-15$~\kms; (11) the velocity of the red-shifted spike; (12) the line width of the red-shifted spike; (13) the flux density of the red-shifted spike. (*) The spectrum has been smoothed to have a channel width of 1.01~\kms.}
\end{deluxetable}

\floattable
\begin{deluxetable}{ccccccccc}
\tablecolumns{9}
\rotate
\tabletypesize{\scriptsize}
\centering
\tablewidth{0pc}
\tablecaption{Summary of observational parameters of Si$^{34}$S (1--0), $^{29}$SiS (1--0), and SiS (2--1) at the Effelsberg-100 m telescope.\label{Tab:obs2}}
\tablehead{
\colhead{Frequency}&\colhead{Transition}&\colhead{$E_{\rm u}/k$}&\colhead{Epoch}      &\colhead{Ch.width}&\colhead{int.time}& \colhead{$v_{\rm s}$} & \colhead{$v_{\rm exp}$}&\colhead{$\int S_{\nu}{\rm d}v$}  \\
\colhead{(MHz)}      &   \colhead{}        &\colhead{(K)}        &  \colhead{}             & \colhead{(\kms)} & \colhead{(min)} & \colhead{(\kms)}  &  \colhead{(\kms)} & \colhead{(Jy~\kms)}  \\          
  \colhead{(1)}      &   \colhead{(2)}     & \colhead{(3)}        &  \colhead{(4)}          & \colhead{(5)}    & \colhead{(6)}   &  \colhead{(7)}    & \colhead{(8)}  & \colhead{(9)}   }
\startdata
17657.7       & Si$^{34}$S (1$\to$0)        &   1         & 2002 Jun. 16      & 1.32     &   803  & $-$25.3$\pm$2.6  &  15.1$\pm$0.4     & 0.157$\pm$0.020               \\
17821.3       & $^{29}$SiS (1$\to$0)        &   1         & 2002 Jun. 16      & 1.32     &   803  & $-$25.9$\pm$2.6  &  12.9$\pm$0.8     & 0.141$\pm$0.022               \\
36309.6       & SiS (2$\to$1)               &   3         & 2001 Sep. 1       & 0.64     &   172 & $-$26.3$\pm$0.6  &  13.9$\pm$0.1     & 12.281$\pm$0.144              \\
\enddata
\tablecomments{(1) the rest frequency of the transition; (2) the name of the transition; (3) the upper energy temperature of the transition; (4) the epoch of the observations carried out; (5) the channel width; (6) the on-source integration time of the observation; (7) the systemic velocity; (8) the expansion velocity which is defined as the half-width at zero power; (9) the integrated intensity.}

\end{deluxetable}

\floattable
\begin{deluxetable}{cccccccccc}
\rotate
\tablecolumns{10}
\tabletypesize{\scriptsize}
\centering
\tablewidth{0pc}
\tablecaption{Observational parameters of interferometric data.\label{Tab:interferometer}}
\tablehead{
\colhead{Line} & \colhead{frequency} & \colhead{Telescope} & \colhead{Config.} & \colhead{Epoch}& \colhead{Weighting} & \colhead{beam}      & \colhead{P.A.}     & \colhead{chan.width} & \colhead{$\sigma$}\\
\colhead{    } & \colhead{(GHz)}     & \colhead{}          & \colhead{} & \colhead{} & \colhead{robust} & \colhead{(\arcsec$\times$\arcsec)} & \colhead{(\arcdeg)}& \colhead{(\kms)} & \colhead{(mJy~beam$^{-1}$)}
}
\startdata
 SiS (1$\to$0)    & 18.1549             & ATCA                & 1.5D, H214, 6C & 2006 Apr. 11 & 2    & 4.42$\times$2.13   & $-$3    & 0.065  & 5.2 \\
 SiS (1$\to$0)    & 18.1549             & ATCA                & 1.5D, H214, 6C & 2006 Apr. 11 & 0    & 3.68$\times$0.38   & $-$2    & 0.065  & 6.2 \\
 SiS (1$\to$0)    & 18.1549             & ATCA                & 1.5D, H214, 6C & 2006 Apr. 11 & $-$2 & 2.62$\times$0.34   & 0       & 0.065  & 30.2 \\
 SiS (1$\to$0)    & 18.1549             & JVLA                &  D             & 2011 Dec. 8   & 0.5  & 3.88$\times$3.34   & $-$10   & 2.064  & 0.4 \\
 SiS (2$\to$1)    & 36.3096             & JVLA                &  D             & 2010 Apr. 26 & 0.5  & 3.33$\times$2.01   & $-$29   & 1.032  & 1.5 \\ 
\enddata
%\tablecomments{(a) Hanning smooth was applied to the JVLA SiS (1–0) data to get ride of the Gibbs ringing effects, which results in the actual spectral resolution a factor of 2 worser.}
\end{deluxetable}

\clearpage

\begin{figure*}[!htbp]
\centering
\includegraphics[width = 0.6 \textwidth]{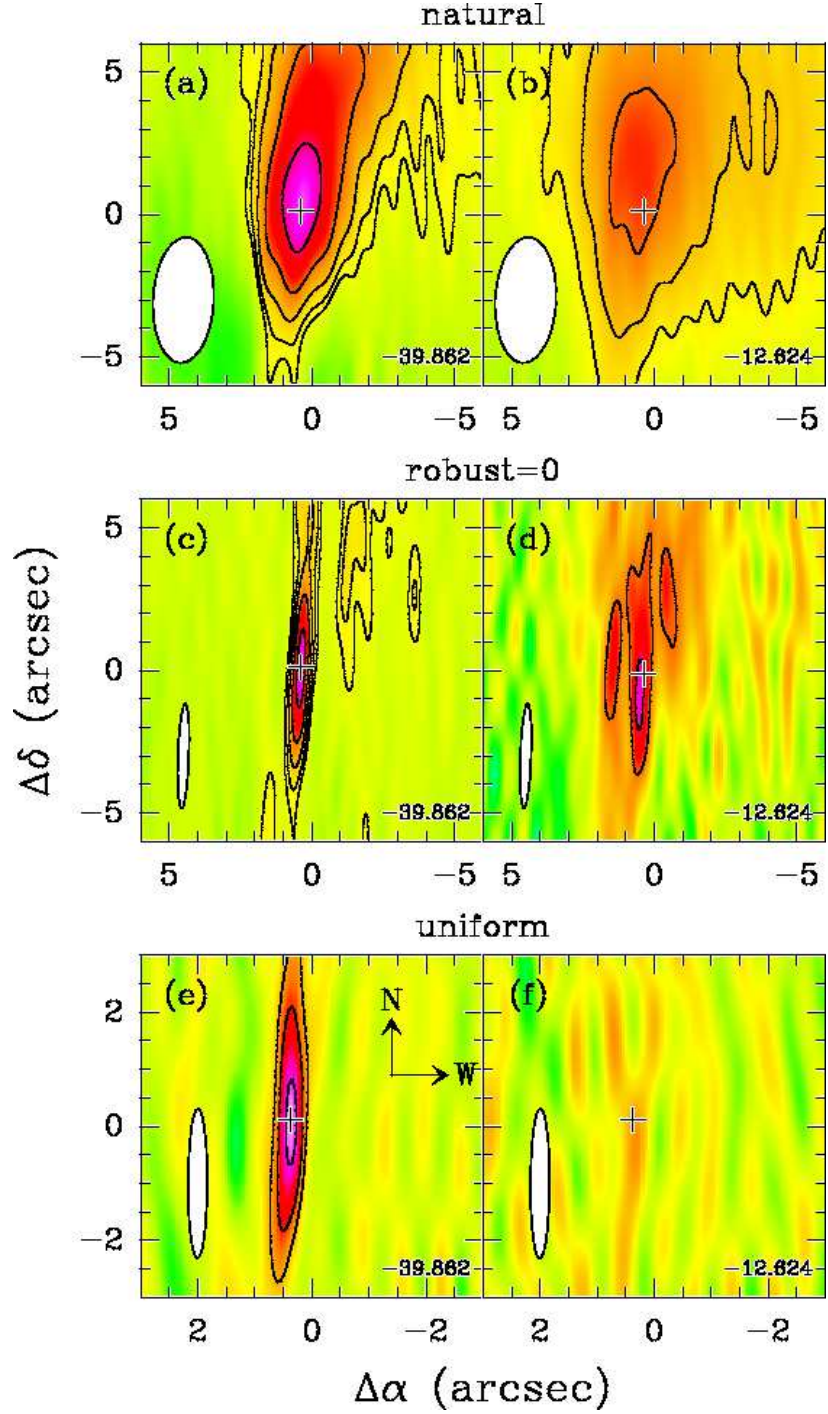}
\caption{{Two ATCA SiS (1$\to$0) channel maps referring to the two spikes, the one near $-$40~\kms\,(left panels) and near $-$13~\kms\,(right panels) .They are cleaned with three different weightings labeled on the top of corresponding panels. The contours start at 5$\sigma$ ($\sigma$=6 Jy~beam$^{-1}$ in Figs.~\ref{Fig:masermap}a---\ref{Fig:masermap}d, and $\sigma$=30 Jy~beam$^{-1}$ in Figs.~\ref{Fig:masermap}e---\ref{Fig:masermap}f), and each contour is twice the previous one. The (0, 0) position in each panel is ($\alpha_{\rm J2000}$, $\delta_{\rm J2000}$)=(09$^{\rm h}$47$^{\rm m}$57$^{\rm s}\!\!$.400, 13\arcdeg16\arcmin43\arcsec$\!\!$.700). The black cross represents the stellar position given by \citet{2012A&A...543A..73M}. The synthesized beam is shown in the lower left of each panel, while velocities are given in the lower right of each panel. } \label{Fig:masermap}}
\end{figure*}

\begin{figure*}[!htbp]
\centering
\includegraphics[width = 0.8 \textwidth]{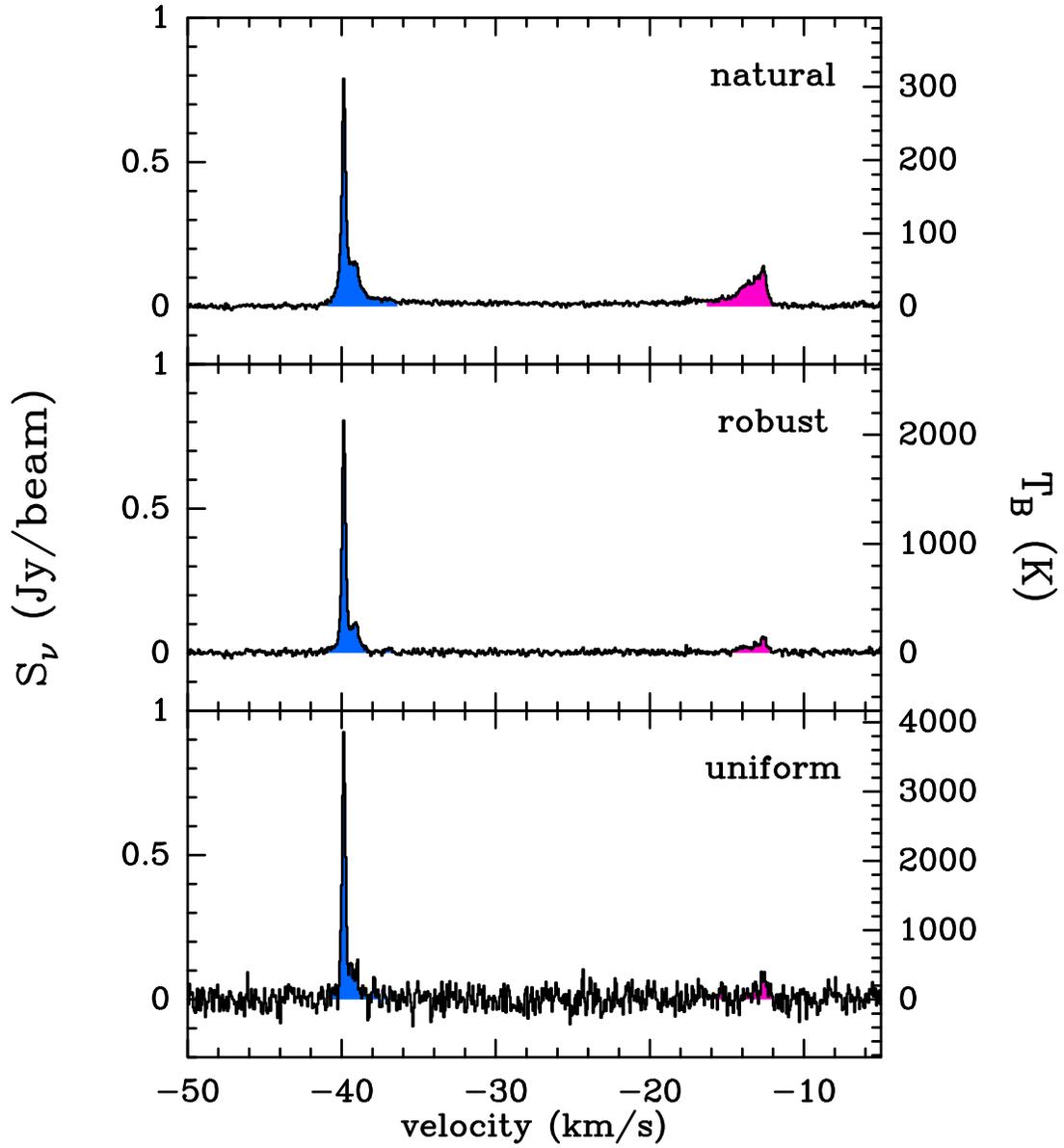}
\caption{{The ATCA SiS (1$\to$0) spectra at the peak of the SiS (1$\to$0) maser (single pixel) with different weightings (natural, robust=0, uniform, indicated in the upper right of each panel) applied in the imaging process. The flux density is shown on the left side, while the brightness temperature is displayed on the right side.} \label{Fig:sis10maser}}
\end{figure*}

\begin{figure*}[!htbp]
\centering
\includegraphics[width = 0.95 \textwidth]{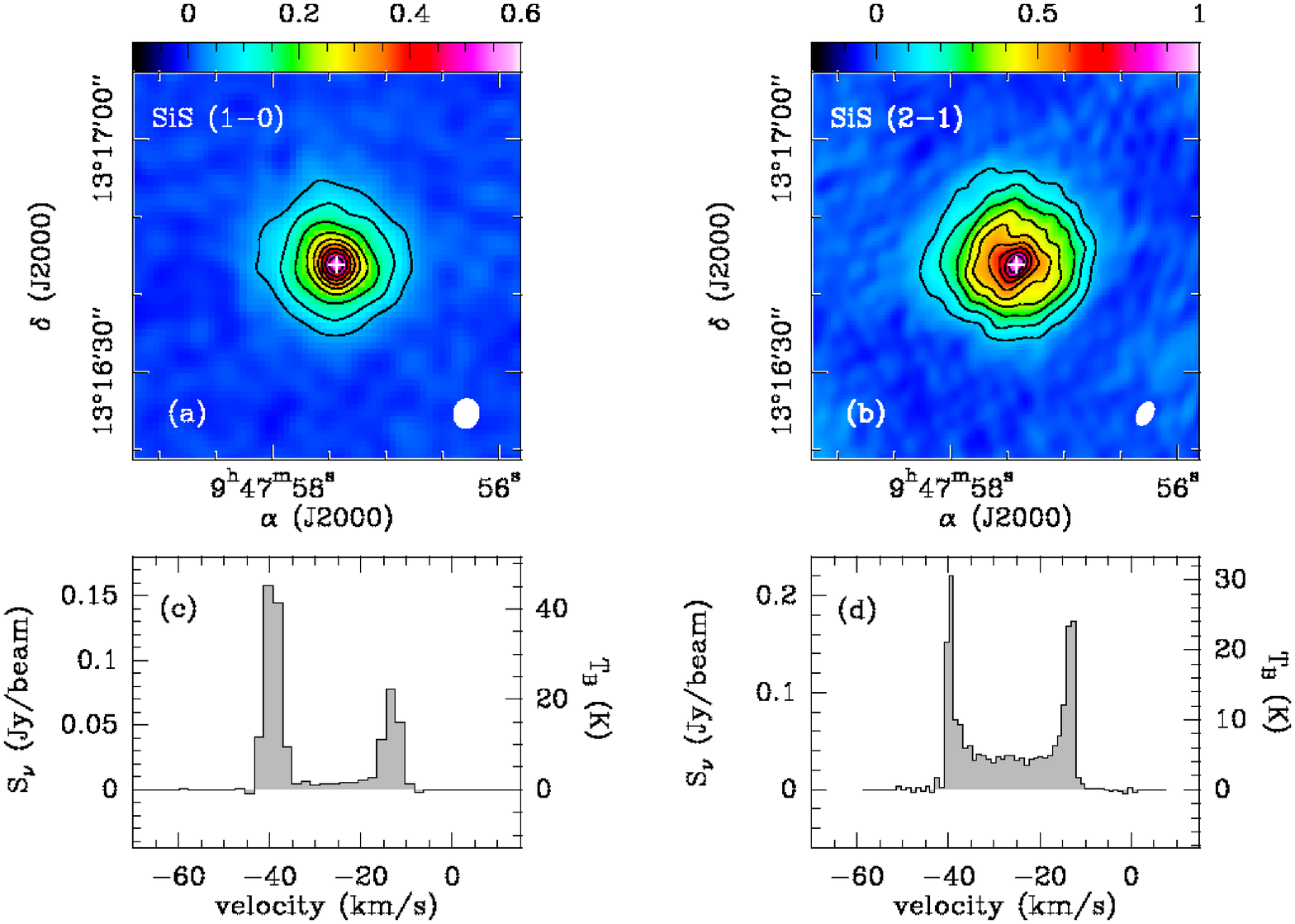}
\caption{{(a) The VLA SiS (1$\to$0) intensity map integrated from $-$38~\kms\, to $-$15~\kms. The contours are 10\%\,to 90\% of the peak intensity (555.1~mJy~beam$^{-1}$~\kms) by steps of 10\%. (b) The VLA SiS (2$\to$1) intensity map integrated from $-$38~\kms\, to $-$15~\kms. The contours are 10\%\,to 90\% of the peak intensity (926.7~mJy~beam$^{-1}$~\kms) by steps of 10\%. In both panels, the color bars represent integrated intensities in units of Jy~beam$^{-1}$~\kms. The white crosses represent the position of the 36.35 GHz continuum source. The synthesized beam is shown in the lower right of each panel. (c) The SiS (1$\to$0) spectrum of the peak in Fig.~\ref{Fig:sismom}a. (d) The SiS (2$\to$1) spectrum from the peak in Fig.~\ref{Fig:sismom}b.} \label{Fig:sismom}}
\end{figure*}

\begin{figure*}[!htbp]
\centering
\includegraphics[width = 0.48 \textwidth]{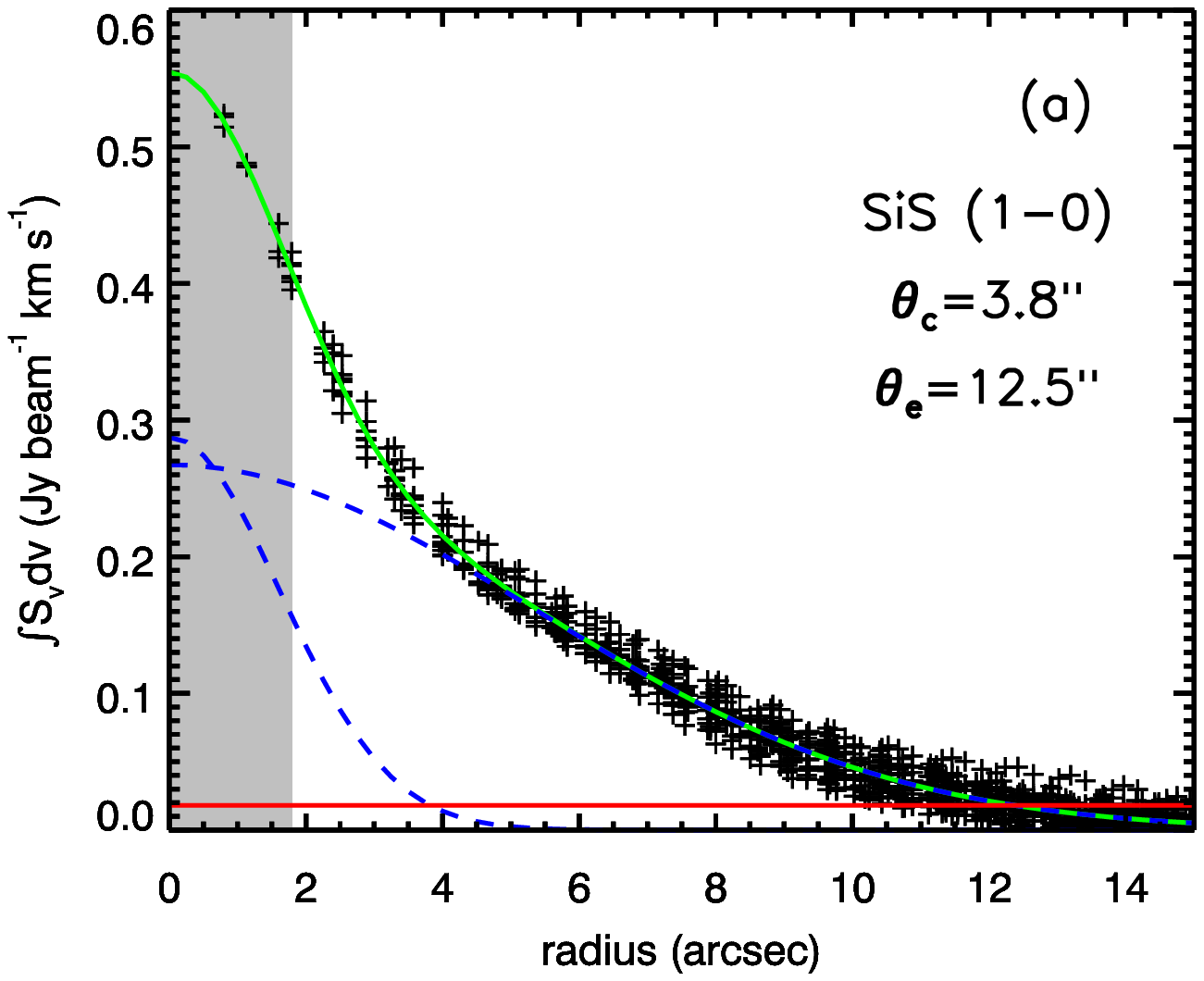}
\includegraphics[width = 0.48 \textwidth]{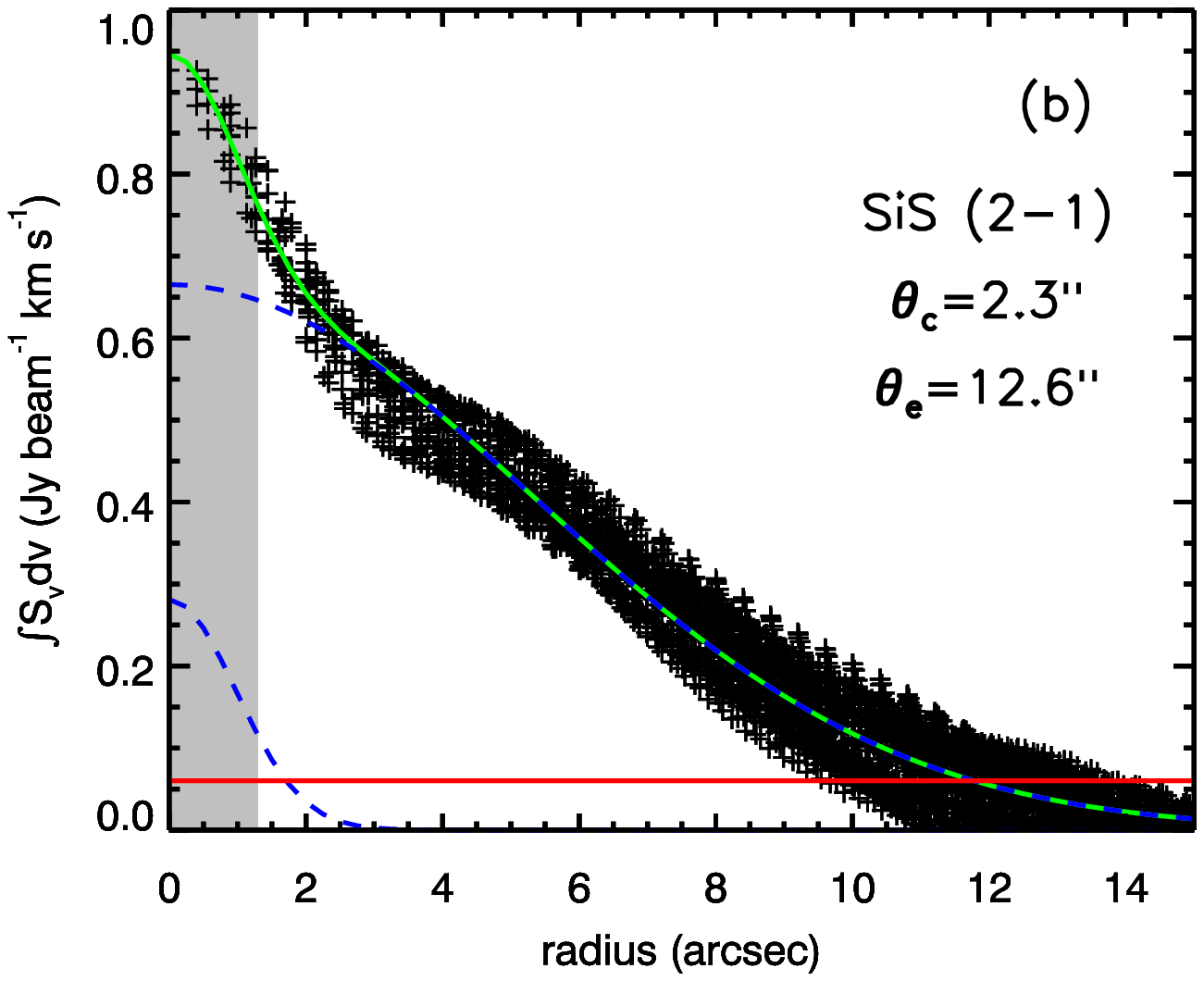}
\caption{{(a) The azimuthally-averaged, velocity-integrated radial intensity profile of SiS (1$\to$0) based on Fig.~\ref{Fig:sismom}a. Each point corresponds to a pixel, and the offset of each pixel is calculated with respect to the pixel with the peak integrated intensity. (b) The same as Fig.~\ref{Fig:intprofile}a but for SiS (2$\to$1) based on Fig.~\ref{Fig:sismom}b. In both panels, the green solid line represents a two-component Gaussian fit. Individual Gaussian components are indicated by the blue dashed lines. The fitted FWHMs for the compact and extended components are indicated by $\theta_{\rm c}$ and $\theta_{\rm e}$. The HPBWs are indicated by the grey shaded areas. The red solid lines mark the detection thresholds ($3\sigma$) which correspond to 18~mJy~beam$^{-1}$~\kms\,for SiS (1$\to$0) and 60~mJy~beam$^{-1}$~\kms\,for SiS (2$\to$1).} \label{Fig:intprofile}}
\end{figure*}

\begin{figure*}[!htbp]
\centering
\includegraphics[width = 0.8 \textwidth]{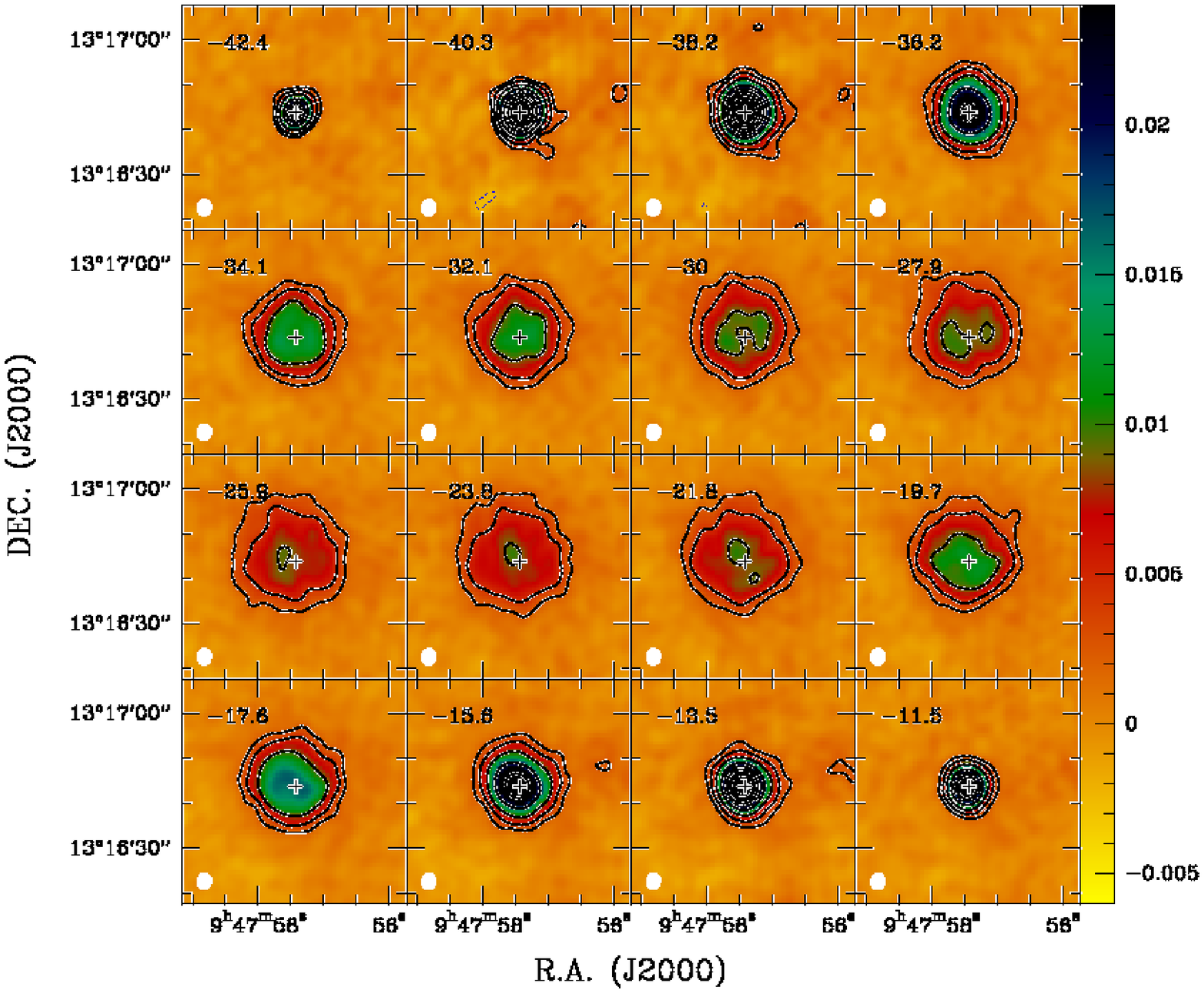}
\caption{{Continuum subtracted VLA channel maps of SiS (1$\to$0) with the velocity labeled in the upper left of each panel. The color bar represents flux densities in units of Jy~beam$^{-1}$. The contours start at 2.25 mJy~beam$^{-1}$ (5$\sigma$), and increase by factors of 2. Dashed contours indicate negative features at a level of $-$2.25 mJy~beam$^{-1}$. The position of the 36.35 GHz continuum source is indicated by a black cross in each panel. The synthesized beam is shown in the lower left of each panel. In this map, a flux density of 0.1~Jy~beam$^{-1}$ corresponds to a brightness temperature of 28.6 K.} \label{Fig:sis10-ch}}
\end{figure*}

\begin{figure*}[!htbp]
\centering
\includegraphics[width = 0.8 \textwidth]{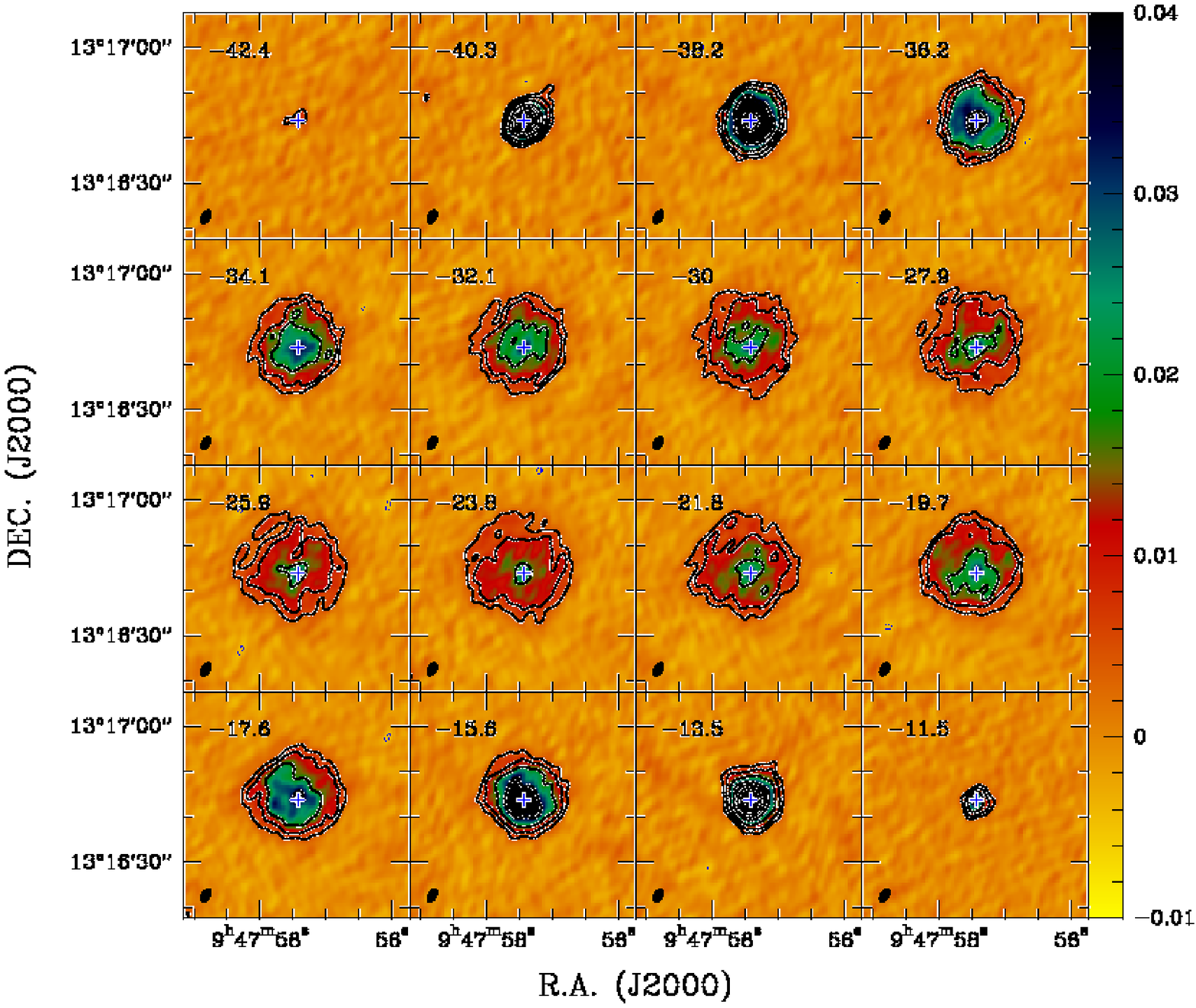}
\caption{{Continuum subtracted VLA channel maps of SiS (2$\to$1) with the velocity labeled in the upper left of each panel. The color bar represents flux densities in units of Jy~beam$^{-1}$. The contours start at 4.5 mJy~beam$^{-1}$ (3$\sigma$), and increase by factors of 2. Dashed contours indicate negative features at a level of $-$4.5 mJy~beam$^{-1}$. The position of the 36.35 GHz continuum source is indicated by a black cross in each panel. The synthesized beam is shown in the lower left of each panel. In this map, a flux density of 0.1~Jy~beam$^{-1}$ corresponds to a brightness temperature of 13.9 K.} \label{Fig:sis21-ch}}
\end{figure*}

\begin{figure*}[!htbp]
\centering
\includegraphics[width = 0.8 \textwidth]{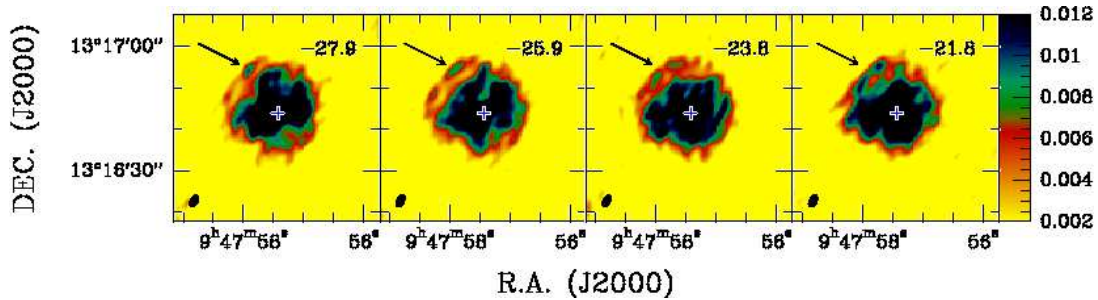}
\caption{{The same as Fig.~\ref{Fig:sis21-ch} but plotted with different color scales. The incomplete shell-like structure is indicated by the black arrow.} \label{Fig:arc}}
\end{figure*}

%\begin{figure*}[!htbp]
%\centering
%\includegraphics[width = 0.48 \textwidth]{fig/sis21profile.eps}
%\caption{{The radial intensity profile of SiS (2$\to$1) at the systemic velocity (based on the $-$26~\kms\,panel of Fig.~\ref{Fig:sis21-ch}). Each point corresponds to a pixel, and the offset of each pixel is calculated with respect to the pixel with the peak integrated intensity. The green solid line represents a two-component Gaussian fit together with individual Gaussian components indicated by the blue dashed lines. The red solid line marks the detection threshold ($3\sigma$) which corresponds to 3.9~mJy~beam$^{-1}$. The red arrows mark four enhanced bumps which appear to be detached shells.} \label{Fig:bump}}
%\end{figure*}

\begin{figure*}[!htbp]
\centering
\includegraphics[width = 0.8 \textwidth]{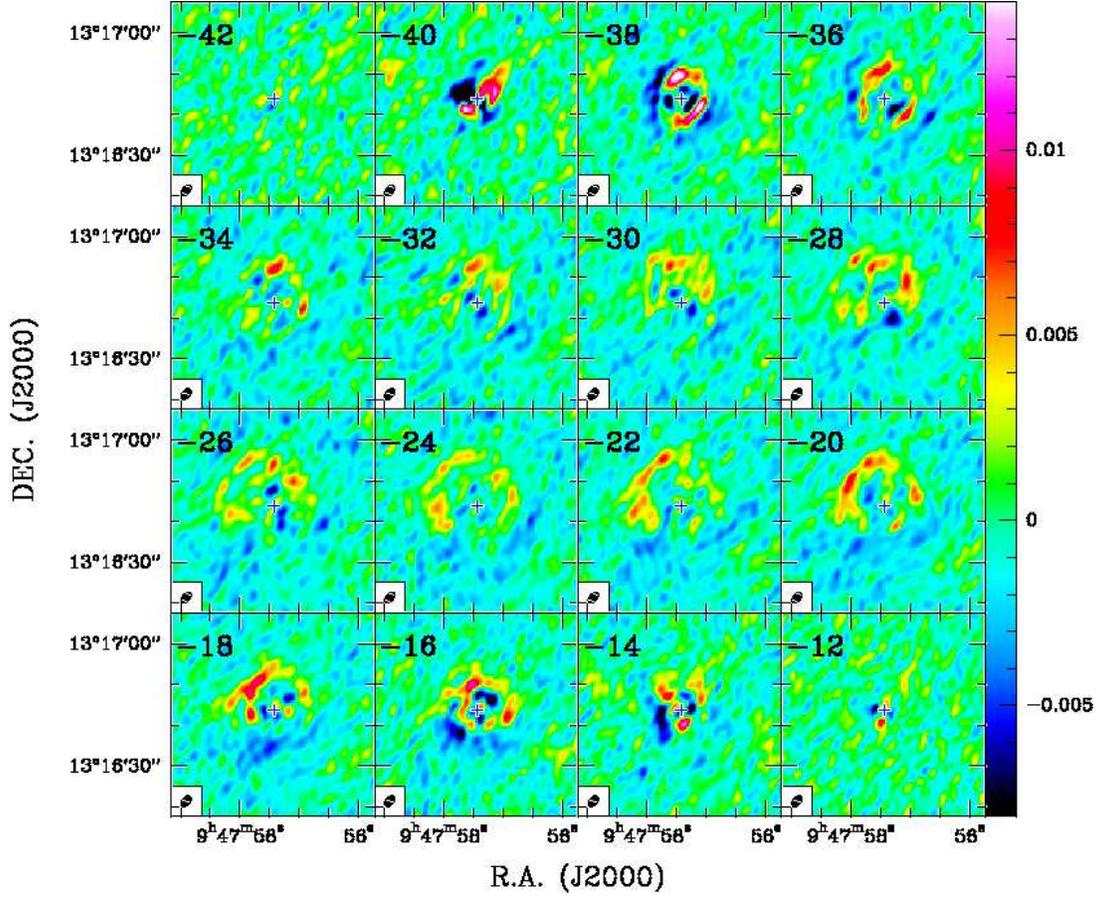}
\caption{{Continuum and spectral background (two components) subtracted VLA channel maps of SiS (2$\to$1) with the velocity labeled in the upper left of each panel. The color bar represents flux densities in units of Jy~beam$^{-1}$. The position of the 36.35 GHz continuum source is indicated by the black cross in each panel. The synthesized beam is shown in the lower left of each panel. In this map, a flux density of 0.01~Jy~beam$^{-1}$ corresponds to a brightness temperature of 1.4 K.} \label{Fig:sis21-resi}}
\end{figure*}

\begin{figure*}[!htbp]
\centering
\includegraphics[width = 0.95 \textwidth]{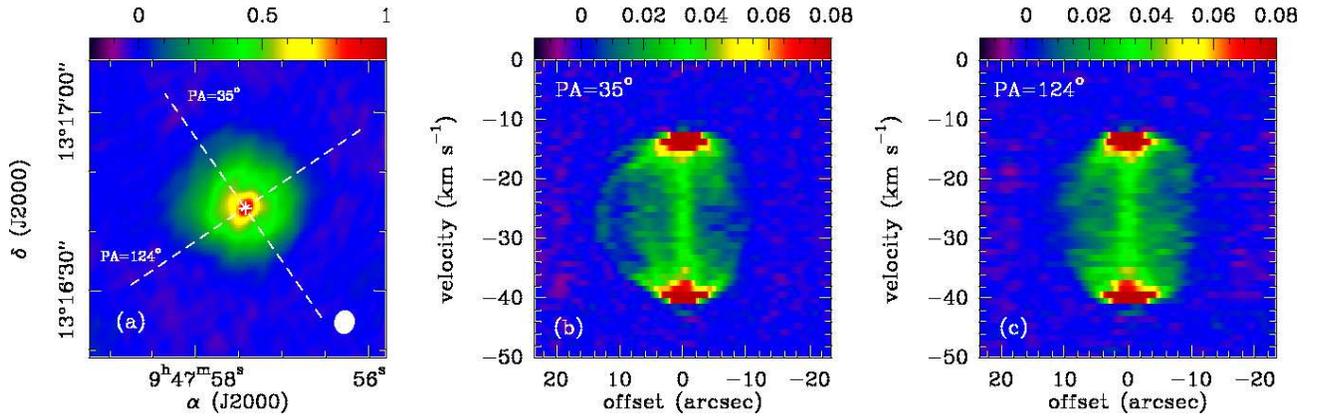}
\caption{{(a) The same as Fig.~\ref{Fig:sismom}b but overlaid with two p-v cuts indicated by two dashed lines. The color bar represents integrated intensities in units of Jy~beam$^{-1}$~\kms. (b) Position-velocity diagram of SiS (2$\to$1) at a PA of 35\arcdeg. (c) Position-velocity diagram of SiS (2$\to$1) at a PA of 124\arcdeg. In Fig.~\ref{Fig:sis21-pv}b and \ref{Fig:sis21-pv}c, the color bars represent flux densities in units of Jy~beam$^{-1}$.} \label{Fig:sis21-pv}}
\end{figure*}

\begin{figure*}[!htbp]
\centering
\includegraphics[width = 0.8 \textwidth]{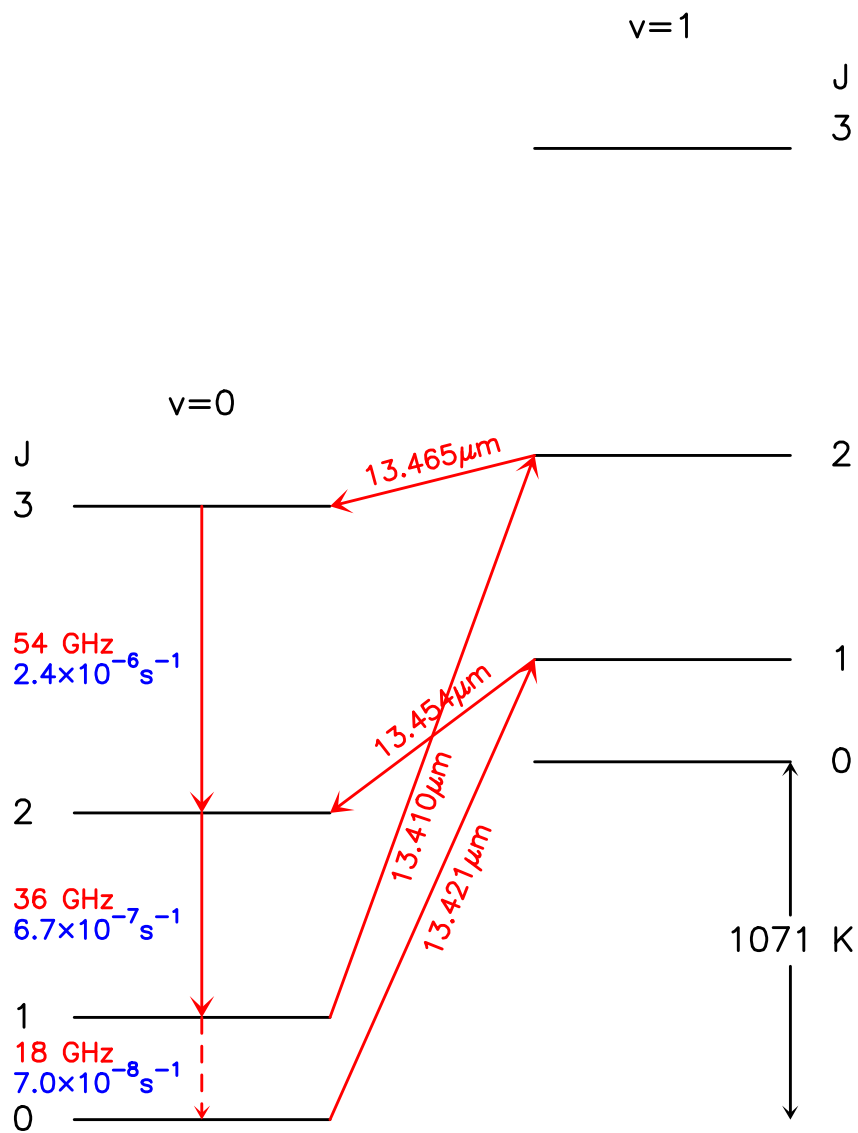}
\caption{{Schematic diagram of energy levels in the two lowest vibrational states of SiS. Frequencies or wavelengths, and the Einstein A coefficients, based on the Cologne Database for Molecular Spectroscopy \citep[CDMS;][]{2005JMoSt.742..215M}, are shown next to corresponding transitions. In this plot, v and $J$ represent vibrational and rotational quantum numbers, respectively.} \label{Fig:energy}}
\end{figure*}

\bibliographystyle{aasjournal}
\bibliography{references}

\clearpage

\appendix
\section{The point spread function of interferometer observations}\label{a.psf}
\begin{figure*}[!htbp]
\centering
\includegraphics[width = 0.8 \textwidth]{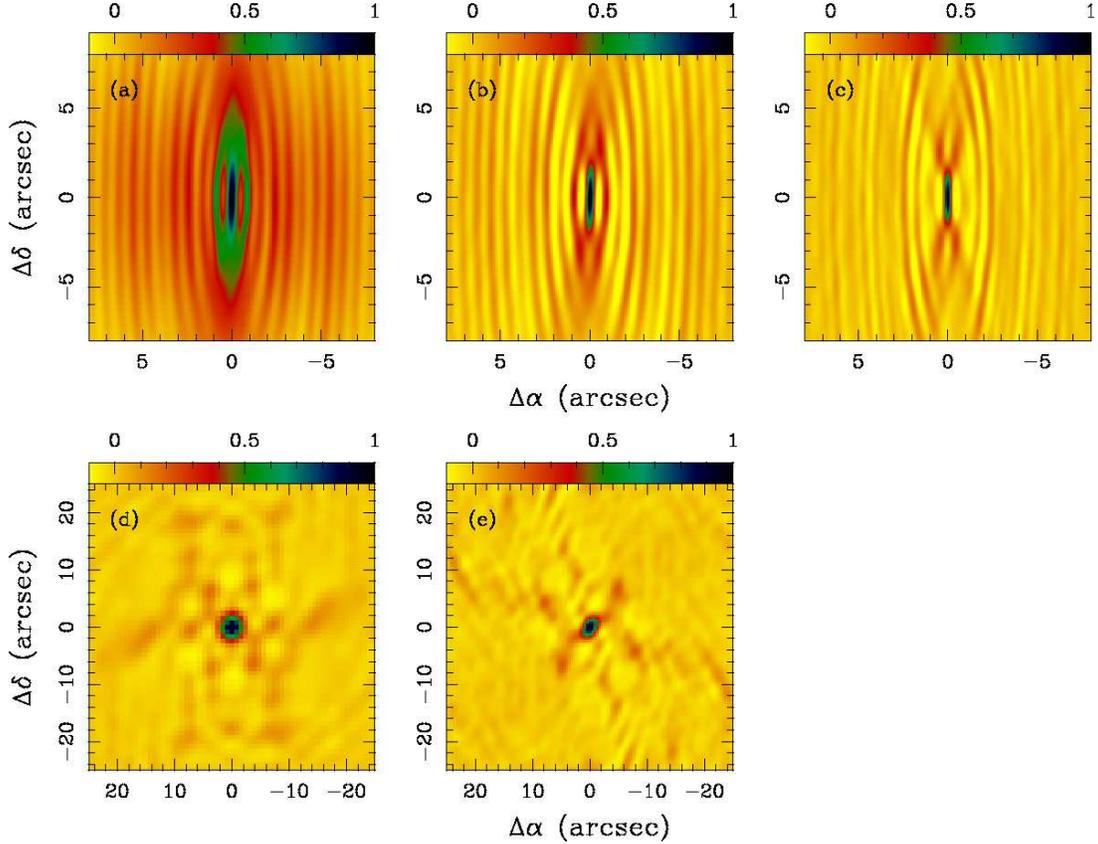}
\caption{{The PSFs of ATCA SiS (1$\to$0) observations with natural (Fig.~\ref{Fig:psf}a), Briggs (robust=0, Fig.~\ref{Fig:psf}b), and uniform (Fig.~\ref{Fig:psf}c) weighting applied. The PSFs of VLA SiS (1$\to$0) and SiS (2$\to$1) observations are shown in Figs.~\ref{Fig:psf}d and \ref{Fig:psf}e, respectively.} \label{Fig:psf}}
\end{figure*}

In order to distinguish artifacts and real structures, the point spread function (PSF) has to be known. Figure~\ref{Fig:psf} gives the PSFs of our ATCA and VLA observations. Figures.~\ref{Fig:psf}a--c show the presence of strong sidelobes which are due to the poor uv coverage. This makes contributions to artifacts in Fig.~\ref{Fig:masermap}. Figure~\ref{Fig:sis21-ch} exhibits elongated northwest-southeast structures in the $-$27.9 and $-$25.9 \kms\,panels. The position angles (PA) of these elongated structures are 113--124\arcdeg, which may be due to the sidelobes seen in Fig.~\ref{Fig:psf}e.

\section{The first and second moment maps of SiS (1$\to$0) and SiS (2$\to$1) emission}\label{sec.mom}
\begin{figure*}[!htbp]
\centering
\includegraphics[width = 0.8 \textwidth]{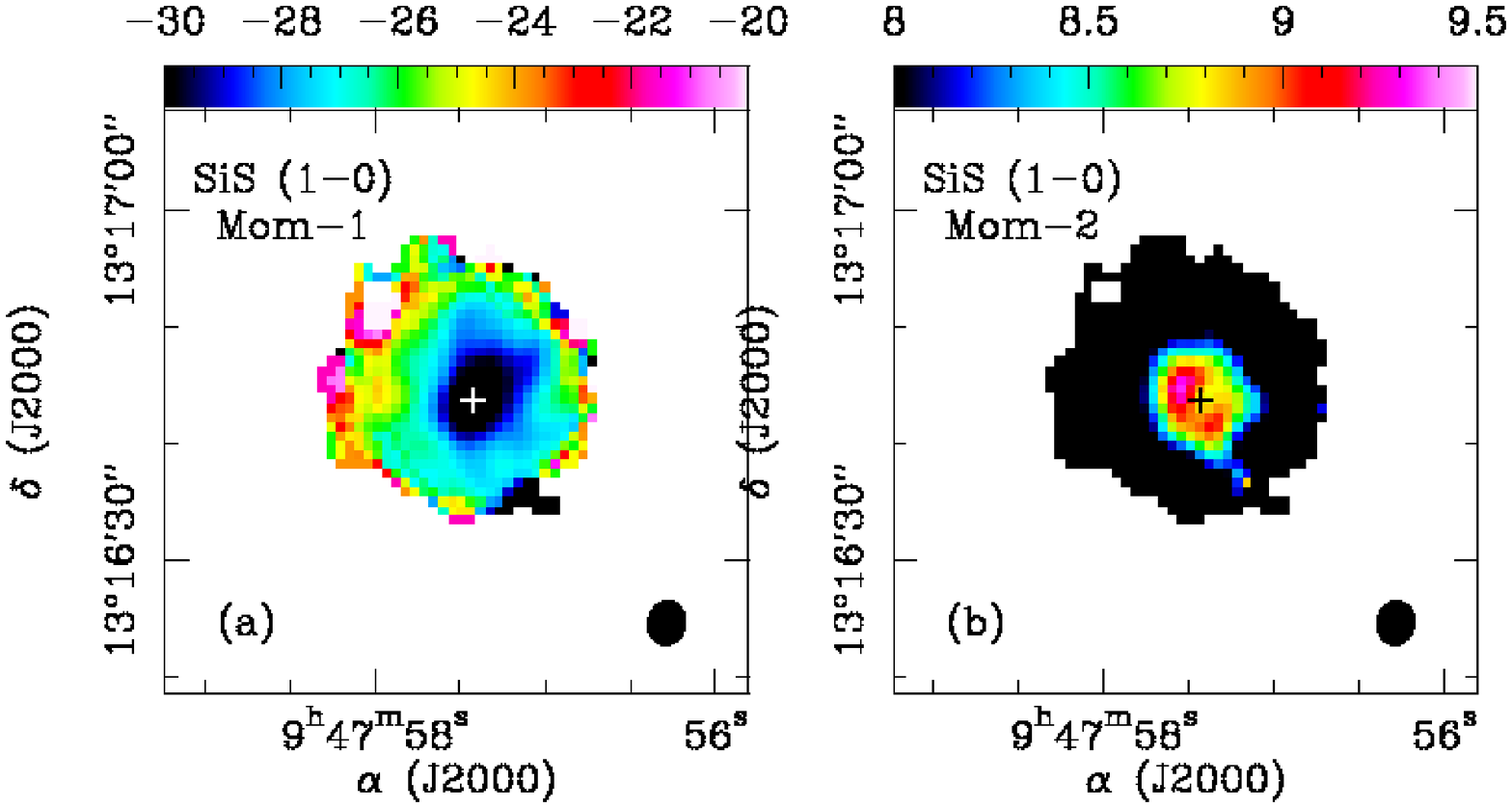}
\includegraphics[width = 0.8 \textwidth]{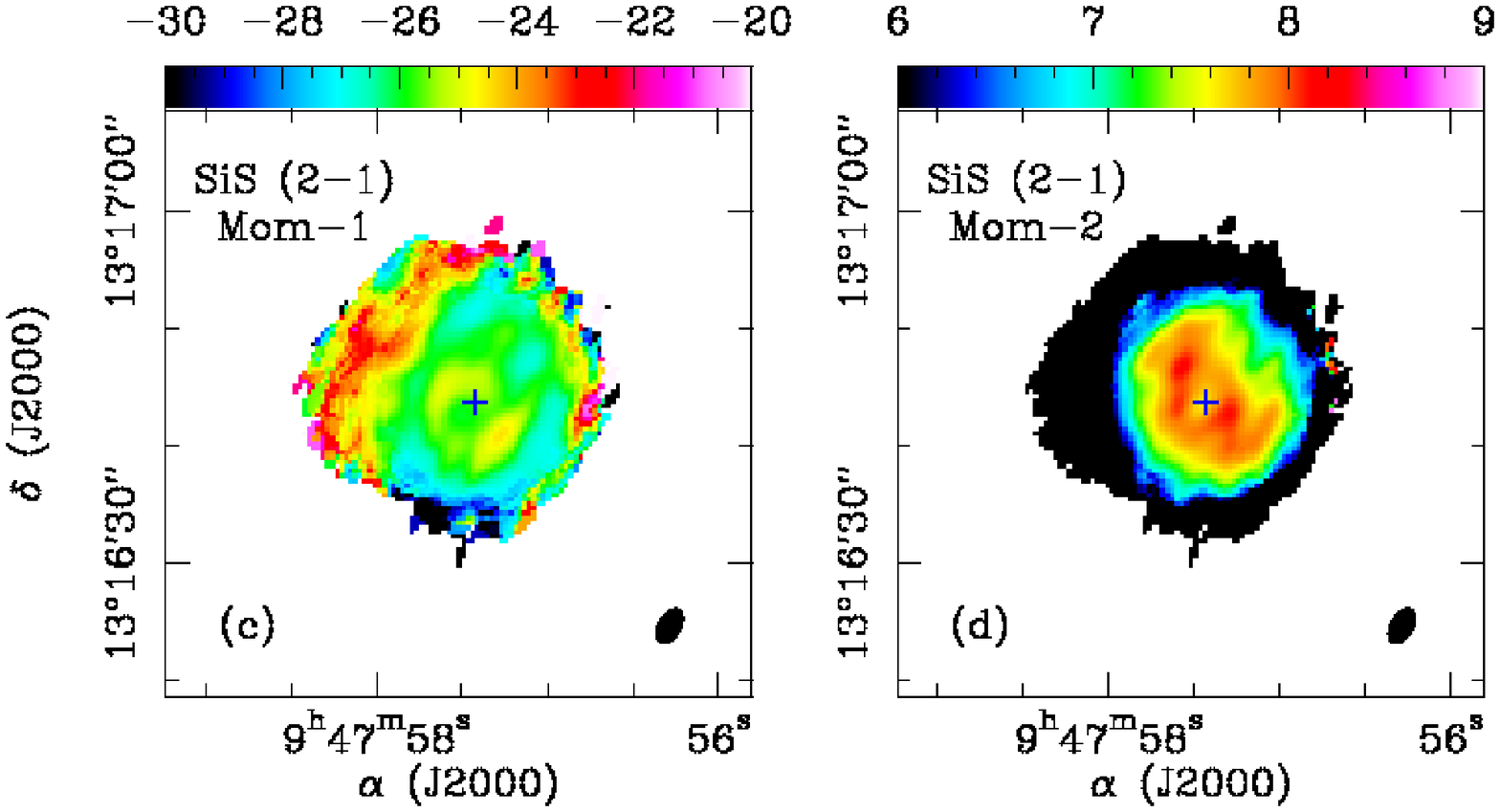}
\caption{{(a) The SiS (1$\to$0) velocity field, clipped at the 3$\sigma$level. The color bar represents the velocities in units of \kms. (b) A map of the SiS (1$\to$0) velocity dispersion, clipped at the 3$\sigma$ level. The color bars represent velocity dispersions in units of \kms. (c) The same as Fig.~\ref{Fig:mom}a but for SiS (2$\to$1). (d) The Same as Fig.~\ref{Fig:mom}b but for SiS (2$\to$1). The synthesized beam is shown in the lower right of each panel.} \label{Fig:mom}}
\end{figure*}

Figure~\ref{Fig:mom} gives the first and second moment maps of SiS (1$\to$0) and SiS (2$\to$1) emission. In the velocity fields (see Fig.~\ref{Fig:mom}a and \ref{Fig:mom}c), we can see red-shifted emission in the north-east, which supports the presence of the incomplete shell proposed in Sect.~\ref{sec.thermal}. The blue-shifted emission in the center of Fig.~\ref{Fig:mom}a is due to the effect of the maser peaking at $-$39.862$\pm$0.065~\kms. Furthermore, there may be shell-like structures in the velocity dispersion maps (see Fig.~\ref{Fig:mom}b and \ref{Fig:mom}d), which is also probably due to episodic mass-loss processes. 

%\section{Convert the flux density to the brightness temperature}
%According to the Rayleigh-Jeans law, the brightness temperature $T_{\rm b}$ can be derived from the flux density of an elliptical Gaussian beam via the formula below:
%\begin{equation}\label{f.tb}
%T_{\rm b} = \frac{\lambda^{2}}{2k}S_{\nu}\Delta\Omega\approx 1359.8913(\frac{S_{\nu}}{\rm Jy~beam^{-1}})(\frac{\lambda}{\rm cm})^{2}\frac{1\arcsec}{\Delta\theta_{\rm maj}}\frac{1\arcsec}{\Delta\theta_{\rm min}}\,{\rm K}\;,
%\end{equation}
%where $T_{\rm b}$ is the brightness temperature, $k$ is the Boltzmann constant, $\Delta\Omega$ is the solid angle occupied by the beam, $S_{\nu}$ is the flux densi%ty in units of Jy~beam$^{-1}$, $\lambda$ is the observed wavelength in units of centimeters, $\Delta\theta_{\rm maj}$ and $\Delta\theta_{\rm min}$ are the major an%d minor axes of the synthesized beam in units of \arcsec.


\end{CJK}
\end{document}